\documentclass[iop,apj,tighten,twocolappendix]{emulateapj}

\usepackage{aas_macros}
\usepackage{apjfonts}
\usepackage{amsmath}
\usepackage{amssymb}
\usepackage[caption=false]{subfig}
\usepackage[plainpages=false, colorlinks=true, anchorcolor=blue, linkcolor=blue, citecolor=blue, bookmarks=false]{hyperref}
\usepackage{graphicx}
\usepackage{comment}
\usepackage{multirow}
\usepackage{url}
\usepackage{color}

\def\apjl{{\it Astrophys. J.} Lett~}

\DeclareMathAlphabet{\mathpzc}{OT1}{pzc}{m}{it}

\begin{document}

\title{Detecting eccentric supermassive black hole binaries with pulsar timing arrays: \\Resolvable source strategies}

\author{
S.~R.~Taylor\altaffilmark{1}\altaffilmark{$\star$}, 
E.~A.~Huerta\altaffilmark{2,3}, 
J.~R.~Gair\altaffilmark{4,5}, \&
S.~T.~McWilliams\altaffilmark{2}.
}

\altaffiltext{$\star$}{\email[email: ]{Stephen.R.Taylor@jpl.nasa.gov}}
\altaffiltext{1}{Jet Propulsion Laboratory, California Institute of Technology, 4800 Oak Grove Drive, Pasadena, CA 91106, USA}
\altaffiltext{2}{Department of Physics and Astronomy, West Virginia University, White Hall, Morgantown, WV 26506, USA}
\altaffiltext{3}{NCSA, University of Illinois at Urbana-Champaign, Illinois 61801, USA}
\altaffiltext{4}{Institute of Astronomy, University of Cambridge, Madingley Rd., Cambridge, CB3 0HA, UK}
\altaffiltext{5}{School of Mathematics, University of Edinburgh, King?s Buildings, Edinburgh EH9 3JZ, United Kingdom}

\date{\today}

%%%%%%%%%%%%%%%%%%%%%%%%%%%%%%%%%%%%%%%%%%%%%%%

\begin{abstract}
The couplings between supermassive black-hole binaries and their environments within galactic nuclei have been well studied as part of the search for solutions to the final parsec problem. The scattering of stars by the binary or the interaction with a circumbinary disk may efficiently drive the system to sub-parsec separations, allowing the binary to enter a regime where the emission of gravitational waves can drive it to merger within a Hubble time. However, these interactions can also affect the orbital parameters of the binary. In particular, they may drive an increase in binary eccentricity which survives until the system's gravitational-wave signal enters the pulsar-timing array band. Therefore, if we can measure the eccentricity from observed signals, we can potentially deduce some of the properties of the binary environment. To this end, we build on previous techniques to present a general Bayesian pipeline with which we can detect and estimate the parameters of an eccentric supermassive black-hole binary system with pulsar-timing arrays. Additionally, we generalize the pulsar-timing array $\mathcal{F}_e$-statistic to eccentric systems, and show that both this statistic and the Bayesian pipeline are robust when studying circular or arbitrarily eccentric systems. We explore how eccentricity influences the detection prospects of single gravitational-wave sources, as well as the detection penalty incurred by employing a circular waveform template to search for eccentric signals, and conclude by identifying important avenues for future study. 
\end{abstract}

%%%%%%%%%%%%%%%%%%%%%%%%%%%%%%%%%%%%%%%%%%%%%%%

\maketitle

%%%%%%%%%%%%%%%%%%%%%%%%%%%%%%%%%%%%%%%%%%%%%%%

\section{Introduction}

The observation of extremely compact objects --- black holes (BHs), neutron stars (NSs), and white dwarfs --- 
and the development of a thorough theoretical understanding of their nature has been one of the triumphs of modern astrophysics~\citep{chandra, mtw,Th300}, but there is still much that we do not understand about these exotic objects. The combination of electromagnetic observations with future detections of gravitational-wave (GW) signals will provide key insights into the nature of compact objects and the role they play in some of the most energetic events in the Universe: gamma-ray bursts, active galactic nuclei, quasars, etc.~\citep{bland,Hughes:2009A, Gebhardt:2000ApJ,Soltan:1982,Peterson:2004,Kormendy:1995,Magorrian:1998,Edo:2013,Berger:2013,Janka:1999,Lee:2007N,Metzger:2012,Piran:2013M,Tanvir:2013}.  Several large-scale collaborations are working to inaugurate the new field of GW astronomy by targeting a wide variety of potential GW sources. These range from the mergers of supermassive black hole binaries (SMBHBs), which may be used by pulsar timing arrays (PTAs) to probe the innermost regions of merging galaxies,
to the coalescence of NS binaries and stellar mass BHs, which encode important information about stellar evolution, galactic nuclei and globular clusters, and are the principle targets for ground-based GW detectors.

In this article, we will focus on a particular type of source that is being targeted by PTAs \citep{fb1990}. PTAs aim to observe GWs in the nanohertz frequency band via the accurate timing of millisecond pulsars. There are three major PTA collaborations --- the European PTA, \citep[EPTA,][]{kc2013}, the North American Nanohertz Observatory for Gravitational-waves \citep[NANOGrav,][]{mclaughlin2013} and the Parkes PTA \citep[PPTA,][]{hobbsPPTA2013} in Australia. These three collaborations also aim to cooperate as the International PTA \citep[IPTA,][]{manchester+2013}. 

The sources of interest in this work are individual SMBHBs during their early inspiral evolution~\citep{rr1995,jb2003,wl2003,sv2010,svv2009}.  Given the nature of these systems, i.e., large orbital separations and small local velocity of the binary components, we can take the compact objects as point-particles without internal dynamics and model the orbital evolution of the system using a post-Newtonian expansion~\citep{peters,barack-cutler,Sesana:2010cq}. Furthermore, these events will be observed at large orbital separations, where the orbital evolution may be more strongly influenced by dynamical interactions with the astrophysical environment rather than GW emission. Hence, the circularizing influence of the latter may be lessened, allowing for quite large orbital eccentricities at the time of detection.

There are several mechanisms that could drive the eccentricity evolution of a SMBHB. For instance, at sub-parsec scales a binary formed by a galactic merger may be embedded in a dense stellar environment. As discussed in~\citet{Sesana:2008S}, if one assumes an isotropic stellar distribution, the interaction of a star and a SMBHB with semi major axis \(a\) can have two possible outcomes. Denoting the semi-major axis of the binary formed by the star and the SMBHB by  \(a_{\star}\), encounters with stars with \(a_{\star} \lesssim a\) tend to circularize the orbit, whereas those with stars with \(a_{\star} \gtrsim a\) tend to increase the eccentricity of the binary. 
In non-isotropic environments, co-rotation of the stellar distribution tends to circularize the binary. Counter-rotating stars tend to extract angular  momentum from the SMBHB, causing the eccentricity to grow~\citep{Sesana:2011M}. Several issues still remain to be explored regarding the evolution of SMBHBs at sub-parsec scales in dense stellar environments, but most models seem to favor a growth in orbital eccentricities before these systems enter the frequency band of PTAs~\citep{Sesana:2010dq,Roeding:2012}. 

Aside from interactions with stars, the dynamical evolution of a SMBHB at sub-parsec orbital separations can also be influenced by the redistribution of energy and angular momentum between the binary and a self-gravitating disc. Consider a gaseous disc co-rotating with a binary, and define \(\lambda \equiv  R_t/a\), where \(R_t\) is the distance of the strongest torque on the binary as measured from the center of mass, and \(a\) is the semi-major axis of the binary. Detailed numerical simulations suggest that the evolution of the orbital eccentricity of a SMBHB embedded in a circumbinary disc is independent of the mass-ratio of the system, but depends sensitively on the location of the inner rim of the disc, \(\lambda\), with respect to the binary's center of mass. For \(2<\lambda<2.5\), it is expected that binaries will converge to a critical eccentricity value \(0.55< e_{*}<0.79\). Binaries with initial eccentricities \(e>e_*\) will undergo a steady decrease in eccentricity, whereas binaries with \(e<e_*\) will experience the opposite behavior. The larger the separation between the rim of the disc and the center of mass of the binary, the longer the system will take to attain \(e_*\)~\citep{Roeding:2011MNRAS}. 

Taking into account these considerations, and the fact that uncertainties about the environments of binaries in realistic galaxy mergers make binary eccentricity a legitimate possibility, we recently introduced a theoretical framework to explore in detail the effect of eccentricity for source detection of potential PTA sources~\citep{hmgt2015}. We now extend that analysis by introducing novel, accurate and efficient pipelines that shed light on the accuracy with which the astrophysical parameters of individually resolved eccentric SMBHBs can be reconstructed. This analysis explores the impact of eccentricity both in terms of source detection and parameter estimation, and presents new statistics to facilitate the analysis. Our approach builds on previous Bayesian \citep{ellis2013,teg2014} and frequentist \citep{babak-sesana-2012,ellis-opt-2012} statistics which have assumed circular gravitational waveform models, and unlike recent studies \citep{zhu2015}, can recover all binary characteristics in addition to providing detection statistics. The latter study defined a frequentist statistic in terms of a harmonic sum over the lowest two harmonics, whereas we proceed from the full GW strain model of an eccentric binary, producing analytic signal models for Bayesian PTA single-source GW searches, and a well-motivated frequentist statistic which fully generalizes that of \citet{babak-sesana-2012} and \citet{ellis-opt-2012}.

This article is laid out as follows. In Section \ref{sec:kepler-problem} we briefly review the orbital trajectories of eccentric binary systems, and how we can analytically solve for the orbital phase at a given time. This is followed in Sec.\ \ref{sec:ecc-time-domain-sec} by a description of the eccentric gravitational waveforms we use, and in Sec.\ \ref{sec:timing-residual-sec} by our model of the perturbations these GWs induce in the times of arrival of radio signals from pulsars. The details of our analysis are provided in Sec.\ \ref{sec:analysis-details}, followed by the results of Bayesian and frequentist signal recoveries from simulated datasets in Sec.\  \ref{sec:results-sec}. In Sec.\ \ref{sec:caveats-future}, we discuss the likely impact of several assumptions that we have made which should be explored further in future studies. We finish with concluding remarks in Sec.\ \ref{sec:conclusion-sec}. In the following we adopt units such that $G=c=1$.

\section{Eccentric binary orbits} \label{sec:kepler-problem}

We briefly review the \textit{Kepler problem} and present the general approach to analytically solve for the orbit of an eccentric binary, reiterating some of the notation and formalism of \citet{yunes-eccentric-2009}, and referring the reader to \citet{goldstein} for a more complete discussion.

\begin{figure}
\centering
\includegraphics[angle=0, width=0.5\textwidth]{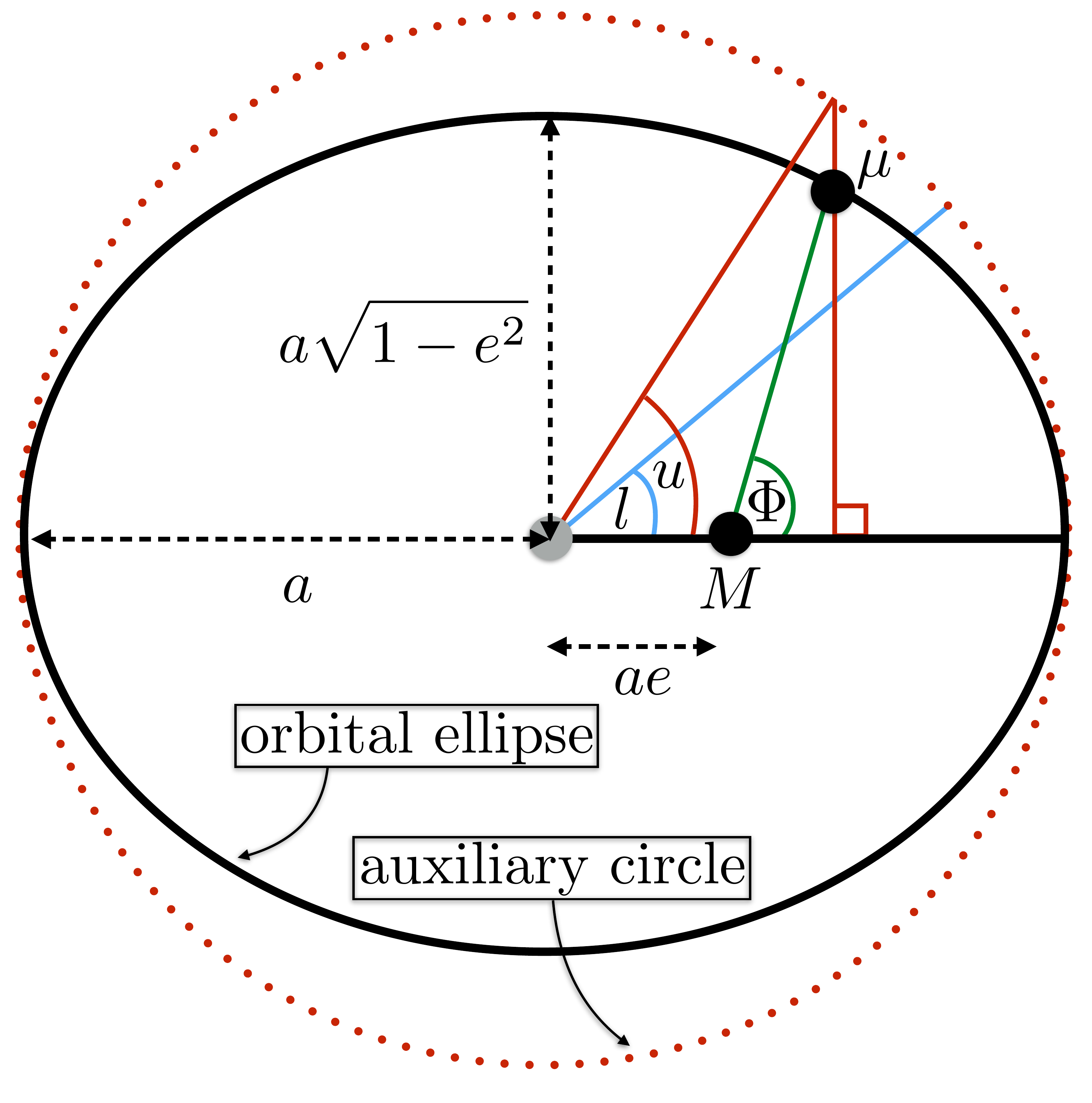}
\caption{A diagram illustrating the relationship between the various angular elements in a binary system with orbital eccentricity $e$, reduced mass $\mu$, and total mass $M$. The semi-major and semi-minor axes are $a$ and $a\sqrt{1-e^2}$, respectively. If we measure the angles from the moment of periapsis, then $\Phi$ is the true anomaly, $l$ is the mean anomaly, and $u$ is the eccentric anomaly. The auxiliary circle has a radius equal to the orbital semi-major axis.}
\label{fig:ecc-angles}
\end{figure}

We consider a binary system with component masses $m_1$ and $m_2$, total mass $M$, and a reduced mass ${\mu=m_1m_2/(m_1+m_2)}$. The separation vector joining the components is defined in terms of the component position vectors by $\vec{r}=\vec{r}_1-\vec{r}_2$, such that $\vec{r}_1=m_2\vec{r}/M$ and $\vec{r}_2=-m_1\vec{r}/M$. Using $(r=|\vec{r}|, \Phi)$ to denote plane polar coordinates for the position of one member of the binary with respect to the other, the Newtonian Keplerian orbital trajectories of two point particles in an eccentric binary system are described by
\begin{equation} 
  r = a(1-e\cos{u}), \label{eq:radial-equation}
\end{equation}
\begin{equation} 
 \omega(t-t_0) = l = u-e\sin{u}, \label{eq:kepler-equation}
\end{equation}
\begin{equation} 
\Phi-\Phi_0 = v \equiv 2\arctan\left[\left(\frac{1+e}{1-e}\right)^{1/2}\tan{\frac{u}{2}}\right],\label{eq:phase-equation}
\end{equation}
where $a$ is the semi-major axis of the orbit, and $0\leq e < 1$ is the eccentricity (of a bound orbit). The \textit{eccentric anomaly}, $u$, is an auxiliary variable with which to parametrize the radial and phase coordinates. Given the average angular frequency (or \textit{mean motion}; $\omega=2\pi/T$, where $T$ is the orbital period) and eccentricity of the orbit, we can solve the transcendental Eq.\ (\ref{eq:kepler-equation}) for $u$ at a given time $t$, where $l=2\pi(t-t_0)/T$ is denoted as the \textit{mean anomaly}. The eccentric anomaly can then be plugged into Eqs.\ (\ref{eq:radial-equation}) and (\ref{eq:phase-equation}) to give the separation and orbital phase (or \textit{true anomaly}; $\Phi-\Phi_0$) at any point along the orbital trajectory. If we assume that time and phase are measured from the moment of periapsis, then the constants of integration $t_0$ and $\Phi_0$ can be set to zero. All of these angular quantities are shown diagrammatically for an example orbital ellipse in Fig.\ \ref{fig:ecc-angles}.

The flux of energy and angular-momentum carried away from the system by GWs depend on the eccentricity and the Keplerian mean orbital frequency, $F$. Once the binary evolution is driven solely by GW emission, these co-evolve as \citep{Peters:1964}
\begin{equation} \label{eq:F-sigma}
\frac{F(e)}{F(e_0)} = \left(\frac{\sigma(e_0)}{\sigma(e)}\right)^{3/2},
\end{equation}
where
\begin{equation} \label{eq:sigma-e}
\sigma(e) = \frac{e^{12/19}}{1-e^2}\left[1+\frac{121}{304}e^2\right]^{870/2299},
\end{equation}
and $e_0$ is defined as the eccentricity of the system at some earlier reference epoch of the binary evolution.

The frequency $F$ can be regarded as the \textit{instantaneous mean orbital frequency}. For GW-dominated orbital evolution, it co-evolves with the eccentricity according to the coupled differential equations \citep{Peters:1964}
\begin{align} \label{eq:dfbydt}
\frac{{\rm d}F}{{\rm d}t} &= \frac{48}{5\pi\mathcal{M}^2}\left(2\pi\mathcal{M}F\right)^{11/3}\frac{1+\frac{73}{24}e^2+\frac{37}{96}e^4}{(1-e^2)^{7/2}},\nonumber\\
\frac{{\rm d}e}{{\rm d}t} &= -\frac{304}{15\mathcal{M}}(2\pi\mathcal{M}F)^{8/3}e\frac{1+\frac{121}{304} e^2}{(1-e^2)^{5/2}},
\end{align}
where $\mathcal{M}=(m_1m_2)^{3/5}/(m_1+m_2)^{1/5}$ is the binary chirp mass. 

Gravitational waveform templates describing the emission from inspiraling binary systems depend on trigonometric functions of the orbital phase. For circular systems the relationship between orbital frequency, time, and phase is simple: we have $\Phi = 2\pi \int F(t) {\rm d}t$, where $F(t)$ is the Keplerian orbital-frequency (half of the dominant quadrupole GW frequency) which evolves according to Eq.\ (\ref{eq:dfbydt}) with $e=0$. However the situation is rather more complicated for eccentric systems. The phase is related via an arctangent to the eccentric anomaly, which is then related to the mean anomaly (and thus the mean angular frequency $\omega=2\pi F$) via a transcendental equation. The so-called \textit{Kepler problem} refers to the historical difficulty in finding solutions to the transcendental equation in Eq.\ (\ref{eq:kepler-equation}) and thus being able to express the orbital phase in terms of the mean anomaly. We do so using the well known Fourier analysis of the Kepler problem. For full details of the calculation see \citet{watson1995treatise}. 
 Using elementary properties of elliptic curves and Bessel functions, the results are
\begin{align} 
\cos\Phi &= -e + \frac{2}{e}(1-e^2)\sum_{n=1}^{\infty}J_n(ne)\cos{(nl)}, \label{eq:cphi-bessel}\\
\sin\Phi &= (1-e^2)^{1/2}\sum_{n=1}^{\infty}\left[J_{n-1}(ne) - J_{n+1}(ne)\right]\sin{(nl)}. \label{eq:sphi-bessel-b}
\end{align}

\begin{figure}[!t]
\centering
\hspace{-0.18in}\includegraphics[angle=0, width=0.5\textwidth]{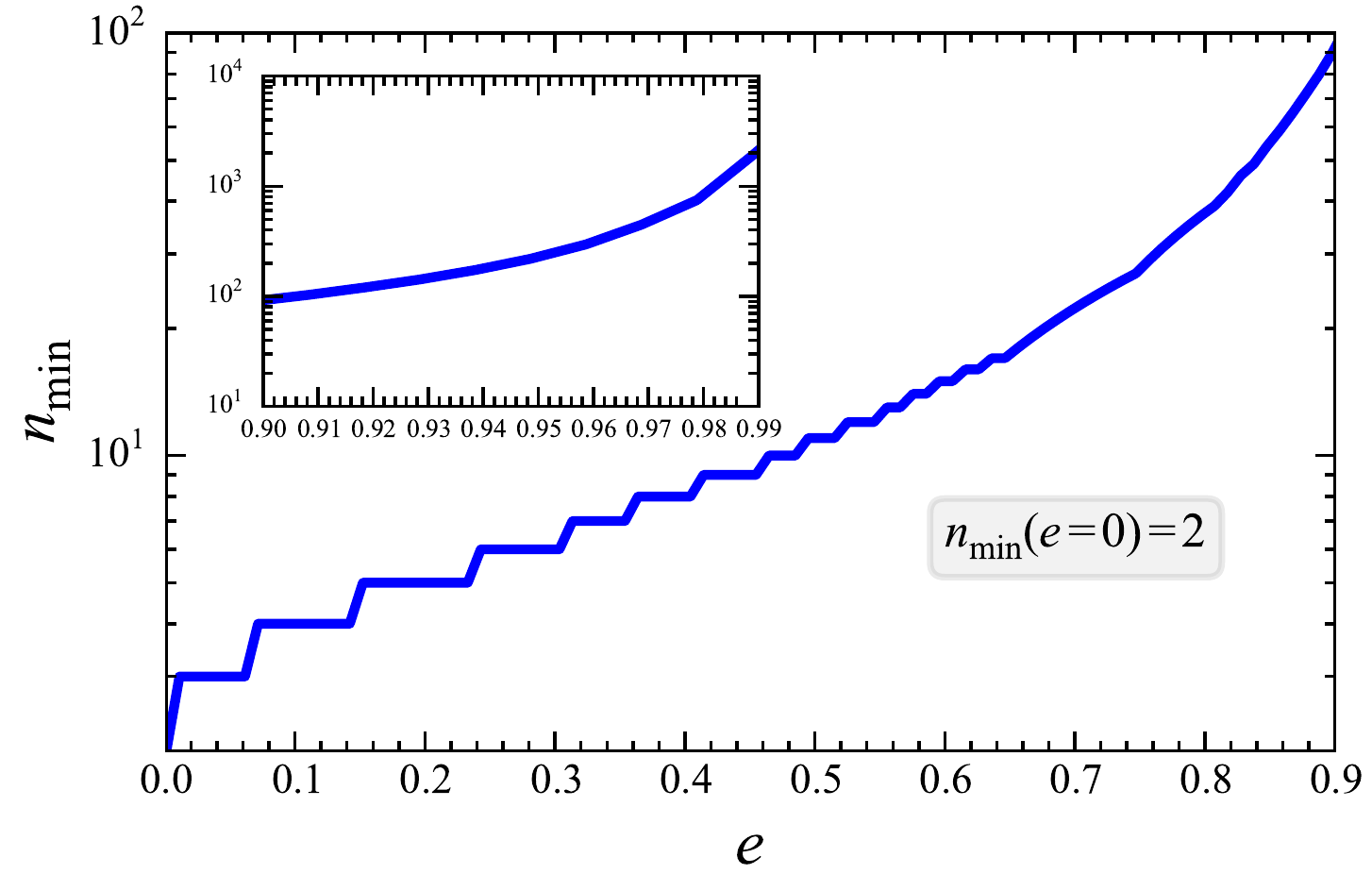}
\caption{The minimum number of harmonics required for the Fourier solution of $\cos\Phi$ as a function of $l$ (mean anomaly) to maintain accuracy with the numerical solution. We demand that the overlap of the Fourier solution and numerical solution, as determined by the normalized scalar product of the two solution vectors, is $> 99.999\%$ over $2\pi$ of mean anomaly.}
\label{fig:minharms}
\end{figure}

With these trigonometric functions of the orbital phase, we can now construct gravitational waveforms for eccentric inspiraling binary systems in terms of the mean orbital frequency. Equations (\ref{eq:cphi-bessel}) and (\ref{eq:sphi-bessel-b}) can be immediately used to construct these waveforms. However, by setting a required tolerance on the accuracy of $\sin\Phi$ and $\cos\Phi$ for a given eccentricity, we can truncate the infinite summations \citep{pierro2001,yunes-eccentric-2009} to accelerate calculations. 
We investigate the minimum number of terms required for the Fourier series expansion of $\cos\Phi$ in Eq.\ (\ref{eq:cphi-bessel}) to maintain accuracy with the exact numerical solution of Eqs.\ (\ref{eq:radial-equation})-(\ref{eq:phase-equation}), by demanding that the error in the two solutions (determined by the normalized scalar product between the two solution vectors) is less than $0.001\%$ over $2\pi$ of mean-anomaly. The results are shown in Fig.\ \ref{fig:minharms}, where we see that $\lesssim 100$ terms in the summation are necessary to maintain accuracy up to $e=0.9$, however the required number of terms dramatically increases beyond $0.9$, exceeding $10^3$ at $e=0.99$. 

Although systems with high residual eccentricity ($> 0.9$) in the sub-parsec GW inspiral regime may exist, they are by no means expected to be common. Unequal mass systems with $q\lesssim 0.25$ may retain $e>0.9$ into the PTA band \citep{Khan}, but we are unlikely to detect their weaker gravitational-wave emission, so we focus here on the more probable case of a detectable signal from a comparable mass binary. Comparable mass binaries in isolated galaxy simulations exhibit $e<0.95$ when they transition from stellar hardening to gravitational-wave-dominated evolution, although preliminary merger simulations can produce binaries with larger eccentricity \citep{Vasiliev}.  However, since gravitational-wave emission is well known to decrease eccentricity \citep{peters}, we believe the assumption that most detectable systems will likely have $e<0.9$ in the PTA band is astrophysically well-motivated, in addition to simplifying things computationally. Ultimately, the range of orbital separations at which the transition between stellar hardening and radiation-reaction occurs in real galaxies is a matter of debate (along with the range of possible binary eccentricities in the PTA band), and may only be resolved with pulsar-timing measurements. Hence, in the following we restrict our attention to systems with eccentricity below $0.9$, in which regime highly accurate waveforms require the inclusion of fewer than $100$ Fourier terms.

\section{Eccentric time-domain waveforms} \label{sec:ecc-time-domain-sec}

In the transverse-traceless gauge the GW-tensor can be written as a linear superposition of ``plus'' and ``cross'' polarization modes, with associated polarization-amplitudes, $h_{\{+,\times\}}$, and basis-tensors, $e_{ab}^{\{+,\times\}}(\hat\Omega)$, such that
\begin{equation}
h_{ab}(t,\hat\Omega) = h_+(t)e^+_{ab}(\hat\Omega) + h_\times(t)e^\times_{ab}(\hat\Omega),
\end{equation}
where $\hat\Omega$ is defined as the direction of GW propagation. 

We employ the Peters-Mathews waveforms \citep{peters} given by \citet{barack-cutler}, which make use of the Fourier analysis of the Kepler problem to give the following analytic expressions for $h_+$ and $h_\times$:
\begin{align} \label{eq:hpluscross-sum}
h_+(t) =& \sum_n -(1+\cos^2\iota)[a_n\cos(2\gamma)-b_n\sin(2\gamma)] \nonumber\\
&+(1-\cos^2\iota)c_n, \nonumber\\
h_{\times}(t) =& \sum_n 2\cos\iota[b_n\cos(2\gamma)+a_n\sin(2\gamma)],
\end{align}
where
\begin{align} \label{eq:hpluscross-coeffs}
 a_n =&- n\zeta\omega^{2/3}\left[J_{n-2}(ne)-2eJ_{n-1}(ne)+(2/n)J_n(ne)\right. \nonumber\\
&\left.\vphantom{J_{n-2}(ne)-2eJ_{n-1}(ne)+(2/n)J_n(ne)}+2eJ_{n+1}(ne)-J_{n+2}(ne)\right]\cos[nl(t)], \nonumber\\
 b_n =&- n\zeta\omega^{2/3}\sqrt{1-e^2}\left[J_{n-2}(ne)-2J_n(ne)+J_{n+2}(ne)\right]\sin[nl(t)], \nonumber\\
 c_n =&\;2\zeta\omega^{2/3}J_n(ne)\cos[nl(t)].
\end{align}
The amplitude parameter is defined as $\zeta=\mathcal{M}^{5/3}/D_L$, where $D_L$ is the luminosity distance of the binary, and $\omega=2\pi F$. The mean anomaly is $l(t)=l_0 + 2\pi \int_{t_0}^t F(t') {\rm d}t'$ (where $l_0$ is the mean anomaly at $t_0$); $\gamma$ is an azimuthal angle measuring the direction of pericenter with respect to $\hat{x}\equiv (\hat\Omega+\hat{L}\cos\iota)/\sqrt{1-\cos^2\iota}$; and $\iota$ is the binary orbital inclination angle, defined by ${\cos\iota=-\hat{L}\cdot\hat\Omega}$. In the following, $F$ and $\mathcal{M}$ refer to the observed redshifted values, such that $F_r = F(1+z)$ and $\mathcal{M}_r = \mathcal{M}/(1+z)$, where $F_r$ and $\mathcal{M}_r$ are rest frame values, and $z$ is the cosmological redshift of the binary.

An important feature to emphasize here is that eccentric binaries do not radiate monochromatic GWs, but rather emit a spectrum of frequencies which are harmonics of the mean orbital frequency. Given that $J_0(0)=1$ and $J_{n>0}(0)=0$, it is immediately obvious from Eqs.\ (\ref{eq:hpluscross-sum}) and (\ref{eq:hpluscross-coeffs}) that $e=0$ waveforms will only include the $n=2$ harmonic of the binary's mean orbital frequency. This is the usual result that the GW frequency of emission from circular binaries is twice the orbital frequency.

\begin{figure}[!t]
\centering
\includegraphics[angle=0, width=0.5\textwidth]{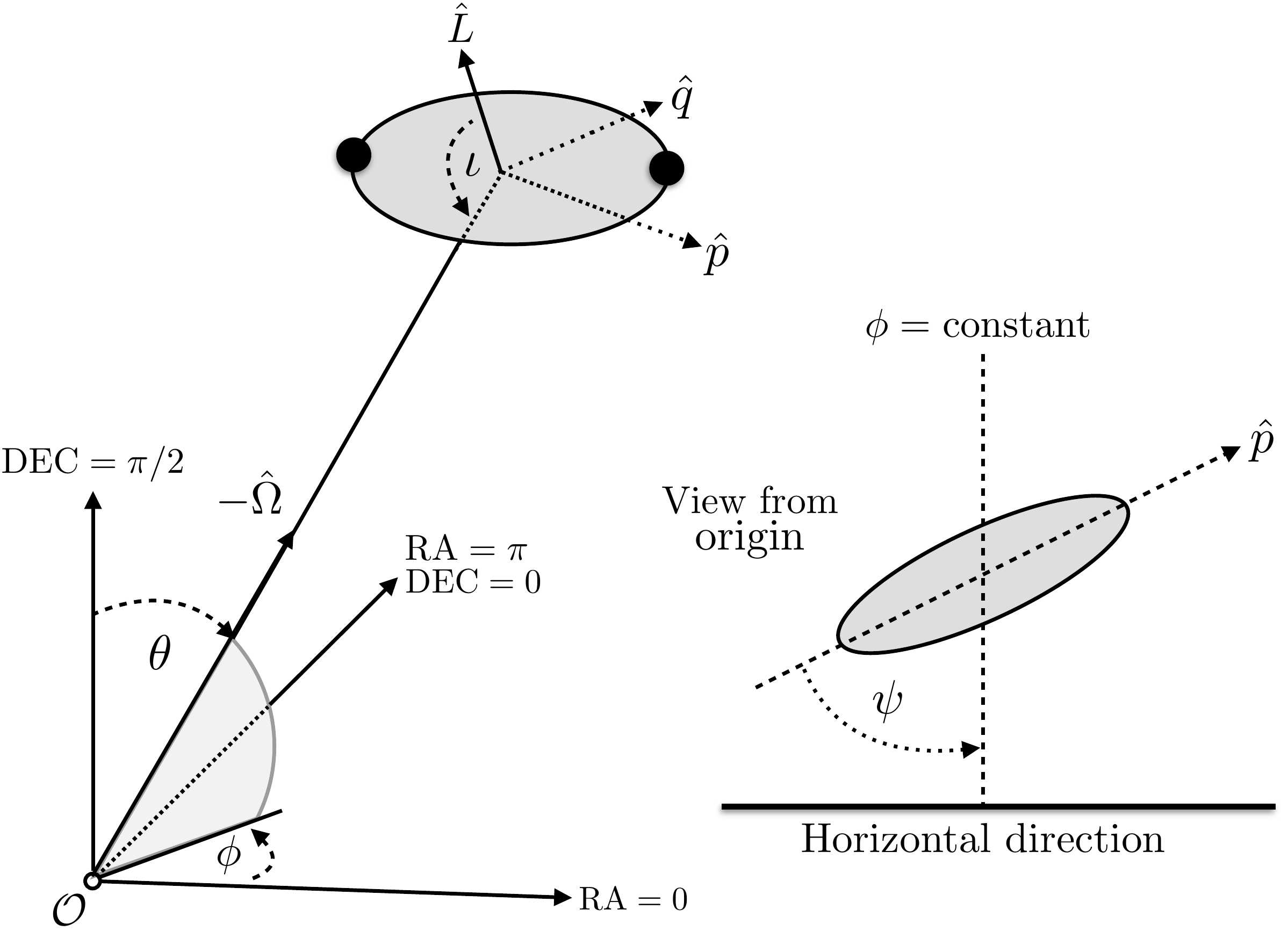}
\caption{A diagram illustrating the geometry of an eccentric SMBHB with respect to the angles of our coordinate system. The unit vector pointing to the binary is $\hat{n}=-\hat\Omega$, with spherical-polar coordinates ${\{\theta=\pi/2-{\rm DEC},\phi={\rm RA}\}}$. The binary orbital inclination angle is defined by $\cos\iota=\hat{L}\cdot\hat{n}$, where $\hat{L}$ is a unit vector pointing along the binary's orbital angular momentum. The GW polarization basis tensors are defined in the plane transverse to the direction of propagation, in terms of the unit vectors $\hat{p}=(\hat{n}\times\hat{L})/|\hat{n}\times\hat{L}|$ and $\hat{q}=\hat{p}\times\hat{n}$, where $\{\hat{n},\hat{p},\hat{q}\}$ define a right-handed basis triad. The vector $\hat{p}$ lies along the major axis of the projected ellipse as seen from the origin of the coordinate system. The GW polarization angle $\psi$ is defined as the angle between $\hat{p}$ and the line of constant azimuth. This diagram is a modified version of Fig.\ $1$ in \citet{apos}.}
\label{fig:detector-angles}
\end{figure}

To construct the polarization basis tensors, we define a right-handed basis triad in terms of $\{\hat{n},\hat{p},\hat{q}\}$, where $\hat{n}=-\hat\Omega$, $\hat{p}=(\hat{n}\times\hat{L})/|\hat{n}\times\hat{L}|$ and $\hat{q}=\hat{p}\times\hat{n}$. The vectors comprising the basis triad are explicitly
\begin{align}
\hat{n} = &\left(\sin\theta\cos\phi, \sin\theta\sin\phi, \cos\theta\right),\\
\hat{p} = &\left(\cos\psi\cos\theta\cos\phi - \sin\psi\sin\phi, \right.\nonumber\\
&\left.\cos\psi\cos\theta\sin\phi + \sin\psi\cos\phi, -\cos\psi\sin\theta\right),\\
\hat{q} = &\left(\sin\psi\cos\theta\cos\phi + \cos\psi\sin\phi,\right. \nonumber\\
&\left.\sin\psi\cos\theta\sin\phi - \cos\psi\cos\phi, -\sin\psi\sin\theta\right),
\end{align}
where $(\theta,\phi) = (\pi/2 - {\rm DEC}, {\rm RA})$ denotes the sky-location of the binary in spherical polar coordinates, and $\psi$ corresponds to the angle between $\hat{p}$ and the line of constant azimuth when the orbit is viewed from the origin of our coordinate system. These angles are shown diagrammatically in Fig.\ \ref{fig:detector-angles}. The vectors $\hat{p}$ and $\hat{q}$ lie in the plane that is transverse to the direction of GW propagation, and are used to construct basis tensors as follows: 
\begin{align} \label{eq:polbasis}
e^+_{ab} = \hat{p}_a\hat{p}_b - \hat{q}_a\hat{q}_b,\\
e^\times_{ab} = \hat{p}_a\hat{q}_b + \hat{q}_a\hat{p}_b.
\end{align}

%%%%%%%%%%

\section{Pulsar timing residuals induced by an eccentric binary} \label{sec:timing-residual-sec}

As a GW transits across the line of sight between a pulsar and the Earth, it creates a perturbation in the space-time metric which causes a change in the proper separation between the Earth and the pulsar. This in turn leads to a shift in the perceived pulsar rotational frequency. The fractional frequency shift of a signal from a pulsar in the direction of unit vector $\hat{u}$, induced by the passage of a single GW propagating in the direction of $\hat\Omega$ is \citep{anholm-2009,brook-flanagan-2011}
\begin{equation}
z(t,\Omega) = \frac{1}{2}\frac{\hat{u}^a\hat{u}^b}{1+\hat\Omega\cdot\hat{u}}\Delta h_{ab}(t,\Omega),
\end{equation}
where $\Delta h_{ab}\equiv h_{ab}(t_e,\hat\Omega) - h_{ab}(t_p,\hat\Omega)$ is the difference in the metric perturbation evaluated at time $t_e$ when the GW passed the solar system barycenter (SSB) and time $t_p$ when the GW passed the pulsar.
From simple geometrical arguments, we can write $t_p = t_e - L(1+\hat\Omega\cdot\hat{u})$, where $L$ is the distance to the pulsar. The integrated effect of this GW-induced {\it redshift} over the total observing time of the pulsar leads to an offset between the expected and the observed pulse TOA: 
\begin{equation}
s(t) = \int_0^tz(t'){\rm d}t'.
\end{equation}
The expected pulse TOA is computed from a deterministic timing model which characterizes a pulsar's astrometric and spin properties. This model is refined over many observations to give an accurate prediction of the pulse arrival times. The difference between the measured TOAs and those predicted by the best-fit deterministic timing-model are the \textit{timing residuals}. In addition to any GW signals, these residuals encode the influence of noise processes and all unmodelled phenomena which affect pulsar TOAs. The pulsar timing residuals induced by a single GW source can be written as
\begin{equation} \label{eq:GWinducedresiduals}
s(t,\hat\Omega) = F^+(\hat\Omega)\Delta s_+(t) + F^\times(\hat\Omega)\Delta s_\times(t),
\end{equation}
where $A=\{+,\times\}$, $\Delta s_A(t) = s_A(t_e) - s_A(t_p)$, with $s_A(t) = \int_0^t h_A(t'){\rm d}t'$, and $F^A(\hat\Omega)$ are {\it antenna pattern response functions}  encoding the geometrical sensitivity of a particular pulsar to a propagating GW, defined as
\begin{equation}
F^A(\hat\Omega)\equiv \frac{1}{2}\frac{\hat{u}^a\hat{u}^b}{1+\hat\Omega\cdot\hat{u}}e^A_{ab}(\hat\Omega),
\end{equation}
and corresponding to the contraction of the pulsar-timing impulse response function with the GW polarization basis tensors.

The form of $s_A(t)$ can be computed analytically by assuming that the binary's mean orbital frequency and eccentricity remain constant over the total timespan of our observations of a given pulsar. More specifically, we must assume no binary evolution over the Earth term timing baseline, ${[t_e, t_e+T]}$, and also the corresponding timing baseline of the pulsar term, ${[t_e - L(1+\hat\Omega\cdot\hat{u}), t_e + T - L(1+\hat\Omega\cdot\hat{u})]}$, where $T$ is $\mathcal{O}(10\;\mathrm{years})$.\footnote{The binary's mean orbital frequency and eccentricity do evolve non-negligibly over the light travel time between the Earth and the pulsar, $\mathcal{O}(1000\;\mathrm{years})$--$\mathcal{O}(10000\;\mathrm{years})$. This effect is easily included in our signal model, however in the rest of this paper we consider only the Earth term.} Therefore, time only appears in the definition of the mean anomaly as a linear parameter, such that $l(t) = l_0 + 2\pi\int^t_{t_0}F(t'){\rm d}t' = l_0 + 2\pi F(t-t_0)$, which allows $\cos[nl(t)]$ and $\sin[nl(t)]$ in Eq.\ (\ref{eq:hpluscross-coeffs}) to be trivially integrated to give the plus/cross residuals:
\begin{align} \label{eq:splusscross-res}
s_+(t) =& \sum_n -(1+\cos^2\iota)[\mathpzc{a}_n\cos(2\gamma)-\mathpzc{b}_n\sin(2\gamma)] \nonumber\\
&+(1-\cos^2\iota)\mathpzc{c}_n, \nonumber\\
s_{\times}(t) =& \sum_n 2\cos\iota[\mathpzc{b}_n\cos(2\gamma)+\mathpzc{a}_n\sin(2\gamma)],
\end{align}
where
\begin{align} \label{eq:splusscross-coeffs}
\mathpzc{a}_n=&\; -\zeta\omega^{-1/3}\left[J_{n-2}(ne)-2eJ_{n-1}(ne)+(2/n)J_n(ne)\right. \nonumber\\
&\left.\vphantom{J_{n-2}(ne)-2eJ_{n-1}(ne)+(2/n)J_n(ne)}+2eJ_{n+1}(ne)-J_{n+2}(ne)\right]\sin[nl(t)] \nonumber\\
=&\; \zeta\omega^{-1/3}\;x_{\mathpzc{a}_n}\!\sin[nl(t)], \nonumber\\
\mathpzc{b}_n =&\;\zeta\omega^{-1/3}\sqrt{1\!-\!e^2}\!\left[J_{n-2}(ne)\!-\!2J_n(ne)\!+\!J_{n+2}(ne)\right]\!\cos[nl(t)] \nonumber\\
=&\;\zeta\omega^{-1/3}\;x_{\mathpzc{b}_n}\!\cos[nl(t)], \nonumber\\
\mathpzc{c}_n =&\;(2/n)\zeta\omega^{-1/3}J_n(ne)\sin[nl(t)] \nonumber\\
=&\;\zeta\omega^{-1/3}\;x_{\mathpzc{c}_n}\!\sin[nl(t)],
\end{align}
and the quantities $\{x_{\mathpzc{a}_n},x_{\mathpzc{b}_n},x_{\mathpzc{c}_n}\}$ are defined for later convenience.

We can now analyze the harmonic content of the variance of the residuals from both plus and cross polarizations, which is computed over one period of binary elliptical motion ($l=\{0,2\pi\}$) and over $\cos\iota,\gamma$. Clearly averaging over a single (or any non-zero integer) period of orbital motion is only an approximation, since our pulsar-timing observations are highly unlikely to span an integer number of orbital periods or GW cycles. Nevertheless we carry out this calculation since it illuminates certain features of the harmonic content of the GW signal from eccentric SMBHBs. We employ the following relations when averaging over the mean anomaly:
\begin{equation}
\int_0^{2\pi}{\rm d}l\; \sin(nl)\cos(n'l) =  0, \;\forall\; n,n', 
\end{equation}
\begin{equation}
\int_0^{2\pi}{\rm d}l\; \sin(nl)\sin(n'l) =  
\begin{cases}
\;0, &\;\text{if}\; n\neq n', \\
\;\pi, &\;\text{if}\; n=n', 
\end{cases}
\end{equation}
where $n,n'\geq 1$, and the last equation is also true for cosine functions. Given that the induced residuals are zero-mean over integers of the binary orbital period, the resulting variance of the residuals is
\begin{equation}
\langle s_A^2\rangle = \zeta^2\omega^{-2/3}\sum_n \langle s_A^2 \rangle_n,
\end{equation}
where 
\begin{align}
\langle s_+^2 \rangle_n &= \frac{7}{15}\left( x_{\mathpzc{a}_n}^2 + x_{\mathpzc{b}_n}^2 \right) + \frac{4}{15}x_{\mathpzc{c}_n}^2, \nonumber\\
\langle s_\times^2 \rangle_n &= \frac{1}{3}\left( x_{\mathpzc{a}_n}^2 + x_{\mathpzc{b}_n}^2 \right).
\end{align}

\begin{figure*} 
\centering
\subfloat{\includegraphics[angle=0, width=0.5\textwidth]{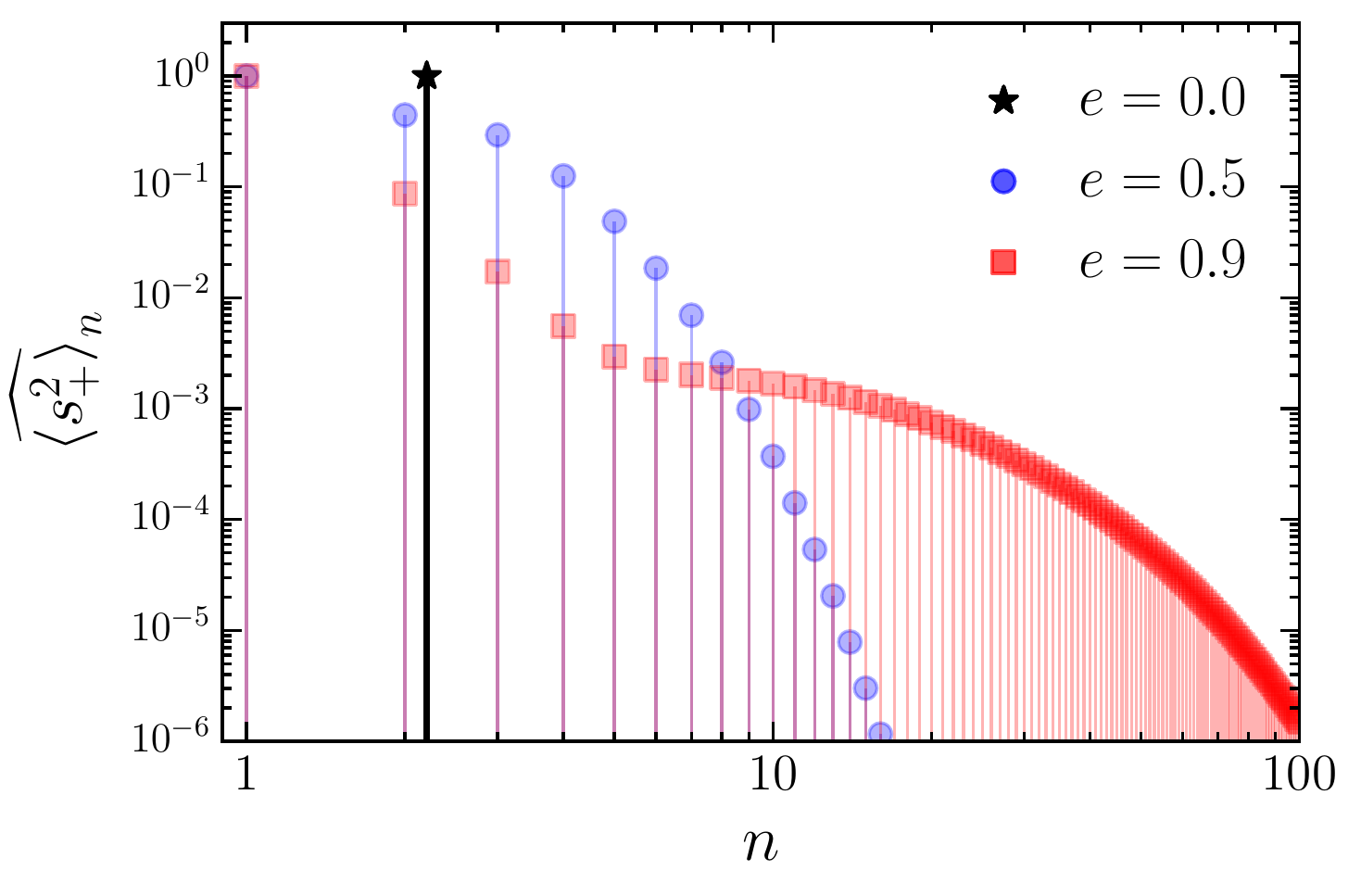}}
\subfloat{\includegraphics[angle=0, width=0.5\textwidth]{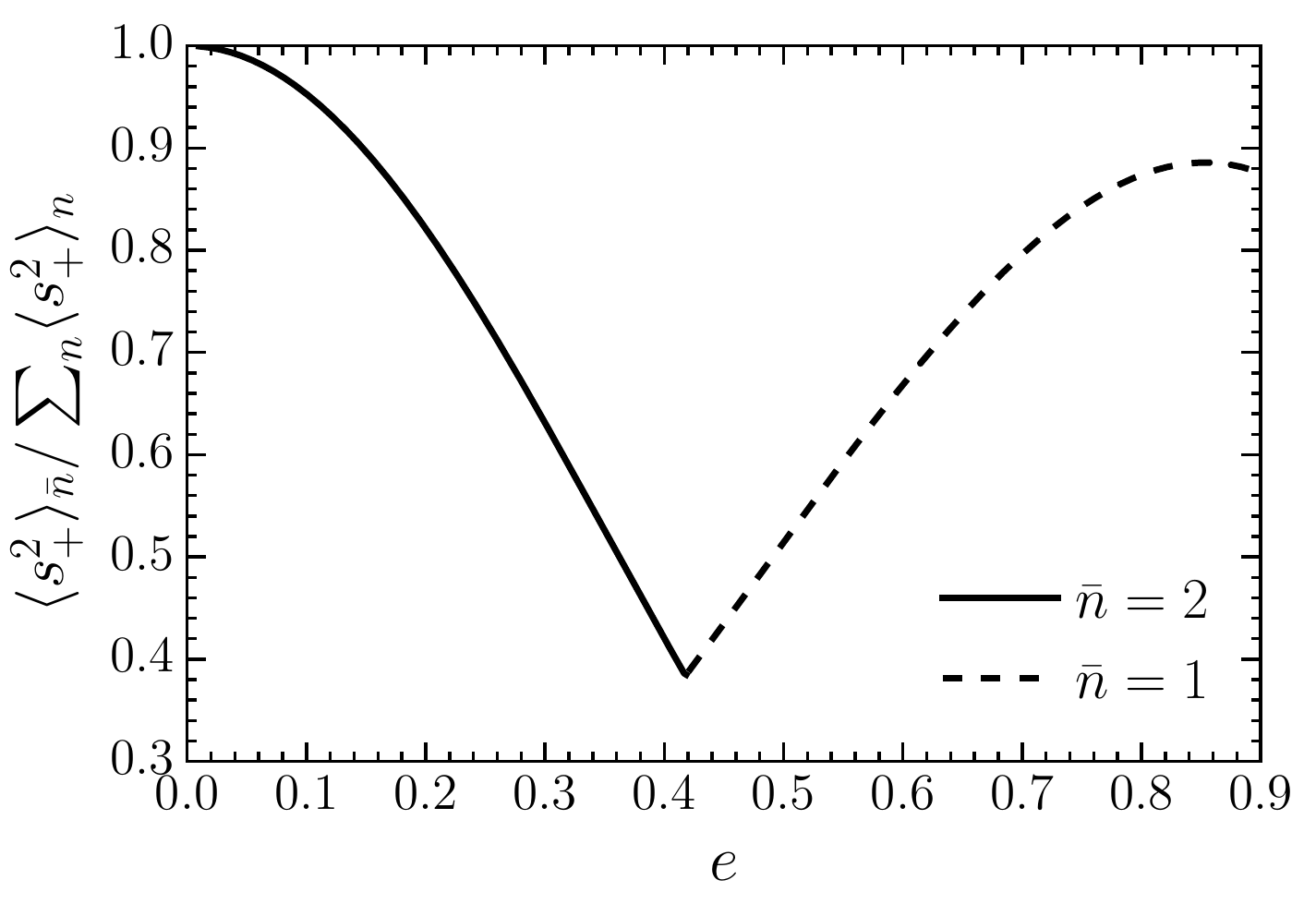}}
\caption{{\it (Left):} The contribution of each harmonic of the orbital frequency to the variance of the plus-component timing residuals. At each eccentricity we normalize the contributions from each harmonic with respect to the maximum contribution. The only contribution for circular binaries is from the second harmonic (black star and line, slightly offset from $n=2$ for ease of viewing). At higher eccentricities ($e=0.5,0.9$) the contribution is spread into a spectrum of higher harmonics, but is dominated by the fundamental harmonic. {\it (Right):} The fraction of the total variance contributed by the dominant harmonic, $\bar{n}$, as a function of eccentricity. As in the left panel, $n$ labels the harmonic of the binary mean orbital frequency. In the range $0\leq e\lesssim 0.4$ the second harmonic dominates, whilst beyond $e\sim 0.4$ the fundamental harmonic dominates the variance of the induced timing residuals.}
\label{fig:PMgne-fig}
\end{figure*}

The value of $\langle s_+^2 \rangle_n$ for several binary eccentricities is shown in the left panel of Fig.\ \ref{fig:PMgne-fig}. At each eccentricity, the contribution of each harmonic to the variance of the residuals is normalized with respect to the largest contribution. In the right panel of Fig.\ \ref{fig:PMgne-fig} we show the fraction of the total variance of the plus-component timing residuals contributed by the dominant harmonic, which switches from $n=2$ in the $0\leq e\lesssim 0.4$ range to $n=1$ beyond $e\sim 0.4$.

For the remainder of this paper we will present results from investigations with the Earth term of the GW-induced timing residuals. The signal model in Eq.\ (\ref{eq:splusscross-res}) is general, and can be used to compute both Earth and pulsar terms, modulo the assumption of binary non-evolution over typical pulsar timing baselines. However, including the pulsar term requires either precise knowledge of the individual pulsar distances, or the distances to be searched or marginalized over \citep{ellis2013,teg2014}. This search over distance brings its own challenges since the likelihood is highly sensitive to small changes in the sampled distance around the true value, and can lead to inefficient sampling. We defer considerations of the pulsar term to future work, but will briefly consider its influence in Sec.\ \ref{sec:caveats-future}. Furthermore, for the most extreme combinations of binary mass, eccentricity, and orbital frequency, the system may exhibit frequency chirping and orbital circularization during typical pulsar-timing observation timespans, rendering the assumption of non-evolution invalid. We explore these issues in Sec.\ \ref{sec:caveats-future} amid suggestions for future directions. 

\begin{figure}[!t]
\centering
\includegraphics[angle=0, width=0.5\textwidth]{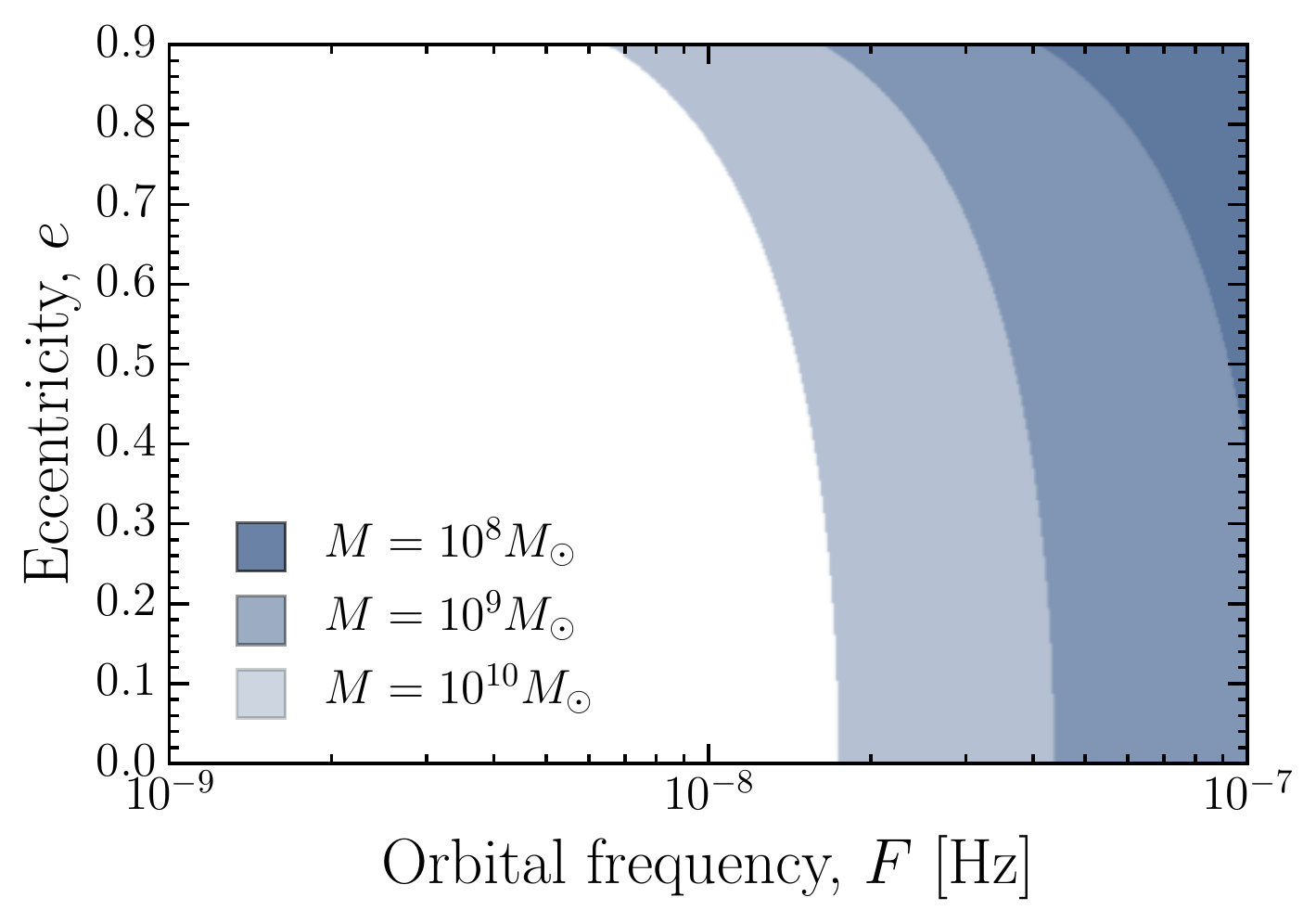}
\caption{Exclusion regions in binary eccentricity and orbital frequency as a function of binary \textit{total mass}, corresponding to parameter combinations where unmodelled evolution of binary pericenter direction causes a bias in orbital frequency recovery which could be resolved by $10$ years of PTA observations, $\Delta f=1/T=3.2$ nHz.}
\label{fig:periapsis-exclusion}
\end{figure}

Related to these two issues are the fact that in general we would also need to consider evolution of the direction of pericenter, $\dot\gamma$, and orbital plane precession from spin-orbit coupling. Evolution of the direction of pericenter can occur even for circular binary systems composed of non-spinning black holes, leading to phase shifts and recovery bias in the orbital frequency if it is not considered. However, as discussed in \citet{Sesana:2010cq}, these factors can be safely ignored over typical PTA observation timespans. In Fig.\ \ref{fig:periapsis-exclusion} we show exclusion regions in $\{M=(m_1+m_2),F,e\}$ parameter space, where pericenter direction evolution leads to a bias in the orbital frequency which is greater than the typical PTA frequency resolution of $1/T$ for a $10$ year observation timespan \citep{Sesana:2010cq}. The excluded regions correspond to systems with very high total mass and eccentricity, and orbital frequencies beyond the region of peak PTA sensitivity. Hence, we ignore this effect here and consider only $\{F,e\}$ evolution in Sec.~\ref{sec:caveats-future}, but information from these additional effects may allow the individual binary component masses, and possibly their spin, to be constrained \citep{Mingarelli:2012dv}. Additionally, these effects are likely to be highly important when tracing the binary evolution back by thousands of years to the pulsar term. 

\section{Simulated datasets and analysis} \label{sec:analysis-details}

For our proof-of-principle study of an eccentric single-source pipeline, we consider two types of PTA datasets. In our \textit{Type I} array, we consider the $36$ pulsars from the IPTA mock data challenge.\footnote{\url{http://www.ipta4gw.org/?page_id=89}} They are timed to $100$ ns precision over a timing baseline of $10$ years, with observations carried out every $4$ weeks. This array is obviously idealized, however the generalization to more realistic observing schedules and pulsar noise properties does not require modifications to our pipeline since it is constructed in the time-domain, and is shielded from Fourier domain spectral leakage caused by red timing noise or irregular sampling. The Bayesian pipeline can be trivially incorporated into a more general pipeline which simultaneously estimates pulsar noise properties and other stochastic signals. The \textit{Type I} datasets will serve as the ideal observing scenario to test for any systematic errors in our signal construction which are separate from observing practicalities, and will also be used for brief analyses of the influence of binary eccentricity on circular- or eccentric-model signal-to-noise ratios (SNRs). 

To emulate more realistic observing schedules and pulsar noise properties, we also construct \textit{Type II} datasets using the actual epochs of observation and noise properties of the $18$ pulsars that were used by the NANOGrav collaboration to place astrophysical constraints on the nanohertz GW background \citep{arz+15a,arz+15b}. These pulsars suffer from irregular sampling, different timing baselines (the longest is $\sim 9$ years), heteroscedastic TOA measurement errors, and, in some cases, intrinsic pulsar spin noise. These \textit{Type II} arrays will be used for our Bayesian studies of the penalties arising from assuming a circular binary model when analyzing data having an eccentric signal, and also when estimating the precision with which current PTAs can estimate binary parameters.

We use the simulation routines within \textit{libstempo},\footnote{\url{http://vallis.github.io/libstempo/}} a python wrapper for the pulsar-timing software package TEMPO2 \citep{tempo2-1,tempo2-2}. For a fiducial source, we are only interested in sensible binary parameters which will illustrate the efficacy of the search pipeline. We follow \citet{ellis2013,teg2014} by considering a source with the following characteristics: $\{\mathcal{M}=10^9M_\odot, F=5\;\text{nHz}, \phi=0.95, \theta = 2.17, \iota=1.57, l_0=0.99, \psi=1.26, \gamma=0.5\}$, and a luminosity distance scaled to meet a required optimal SNR. The definitions of optimal and matched-filtering SNR follow from \citet{finn2001}.

The binary parameter space is searched using a python wrapper \citep{pymultinest} to the nested sampling package \textsc{MultiNest} \citep{mnest-1,mnest-2,mnest-3}, and we have cross-checked our results with a sampler utilizing advanced Markov chain Monte Carlo techniques.\footnote{\url{https://github.com/jellis18/PTMCMCSampler}} The product of these analyses are samples from the posterior probability distribution of the signal parameters space, allowing us to quantify the measurement precision of parameters based on the Bayesian credible regions, and also permitting model selection via computation of competing models' Bayesian evidence. The priors for the signal parameters are as follows: $\log_{10}(\mathcal{M}/M_\odot)\in U[7,10]$, $\log_{10}(D_L/\mathrm{Mpc})\in U[0,4]$, $\log_{10}(F/\mathrm{Hz})\in U[-9.3,-6.0]$, $e\in U[0,0.9]$, $\phi\in U[0,2\pi]$, $\cos\theta\in U[-1,1]$, $\cos\iota\in U[-1,1]$, $\psi\in U[0,\pi]$, $\gamma\in U[0,\pi]$, $l_0\in U[0,2\pi]$. 

Full details of Bayesian inference in the context of PTAs can be found in \citet{vH+09,vHv14,arz+15b}, and for details of how Bayesian searches for continuous GWs are carried out see \citet{ellis2013,teg2014}.

\subsection{Eccentric $\mathcal{F}_e$ statistic} \label{sec:eccFstat-sec}

Equation (\ref{eq:splusscross-res}) provides the appropriate signal model to use when we wish to map out the posterior distribution of the entire signal parameter space, and also if we were to simultaneously search for continuous GW sources in addition to stochastic signal or noise processes. However, we can also construct a fixed-noise frequentist statistic for eccentric binary systems.

We wish to construct a form of the $\mathcal{F}_e$ statistic \citep{babak-sesana-2012,ellis-opt-2012} which can be applied to GW signals from binaries with arbitrary eccentricity. In practice, as in the rest of this paper, we only consider systems with $e\in[0,0.9]$. The $\mathcal{F}_e$ statistic as it is constructed in \citet{ellis-opt-2012} is a maximum-likelihood estimator of the source's sky-location and orbital frequency, and requires that the expression for the induced residuals be rearranged into a form which permits maximization of the likelihood-ratio over the coefficients of a set of time-dependent basis-functions. The likelihood-ratio, $\Lambda$, is defined as the ratio of the likelihood of the data in a model which includes a signal to the noise-only null hypothesis:
\begin{align} \label{eq:likelihood-ratio}
\ln\Lambda =& \ln\left[\frac{\mathcal{L}(\mathbf{s} | \delta\mathbf{t})}{\mathcal{L}(\mathbf{0} | \delta\mathbf{t})}\right] \nonumber\\
=& (\delta\mathbf{t}|\mathbf{s}) - \frac{1}{2}(\mathbf{s}|\mathbf{s}).
\end{align}

We extend the $\mathcal{F}_e$ statistic by rewriting the Earth term residuals (in a single pulsar) given by Eqs.\ (\ref{eq:splusscross-res}) and (\ref{eq:splusscross-coeffs}) as:
\begin{equation} \label{eq:res-basis}
s(t) = \sum_{i=1}^6 \mathpzc{w}_i\mathcal{W}^i,
\end{equation}
where,
\begin{align}
\mathpzc{w}_1 =& \zeta\left[-\left(1+\cos^2\iota\right)\cos(2\gamma )\cos(2 \psi )+2\cos\iota\sin(2\gamma )\sin (2 \psi ) \right],\nonumber\\
\mathpzc{w}_2 =& \zeta\left[\left(1+\cos^2\iota\right) \sin(2\gamma)\cos(2\psi )+2\cos\iota\cos(2\gamma)\sin(2\psi)\right],\nonumber\\
\mathpzc{w}_3 =& \zeta\left[(1-\cos^2\iota)\cos(2\psi)\right],\nonumber\\
\mathpzc{w}_4 =& \zeta\left[\left(1+\cos^2\iota\right)\cos(2\gamma )\sin(2 \psi )+2\cos\iota\sin(2\gamma )\cos(2 \psi ) \right],\nonumber\\
\mathpzc{w}_5 =& \zeta\left[-\left(1+\cos^2\iota\right) \sin(2\gamma)\sin(2\psi )+2\cos\iota\cos(2\gamma)\cos(2\psi)\right],\nonumber\\
\mathpzc{w}_6 =& \zeta\left[-(1-\cos^2\iota)\sin(2\psi)\right],
\end{align}
\begin{align} \label{eq:Fe-basis}
\mathcal{W}^1 &= \tilde{F}^+(\hat\Omega)\omega^{-1/3}\sum_nx_{\mathpzc{a}_n}\sin[n\omega(t-t_0)+nl_0], \nonumber\\
\mathcal{W}^2 &= \tilde{F}^+(\hat\Omega)\omega^{-1/3}\sum_nx_{\mathpzc{b}_n}\cos[n\omega(t-t_0)+nl_0], \nonumber\\
\mathcal{W}^3 &= \tilde{F}^+(\hat\Omega)\omega^{-1/3}\sum_nx_{\mathpzc{c}_n}\sin[n\omega(t-t_0)+nl_0], \nonumber\\
\mathcal{W}^4 &= \tilde{F}^\times(\hat\Omega)\omega^{-1/3}\sum_nx_{\mathpzc{a}_n}\sin[n\omega(t-t_0)+nl_0], \nonumber\\
\mathcal{W}^5 &= \tilde{F}^\times(\hat\Omega)\omega^{-1/3}\sum_nx_{\mathpzc{b}_n}\cos[n\omega(t-t_0)+nl_0], \nonumber\\
\mathcal{W}^6 &= \tilde{F}^\times(\hat\Omega)\omega^{-1/3}\sum_nx_{\mathpzc{c}_n}\sin[n\omega(t-t_0)+nl_0],
\end{align}
and we adapt the number of terms in these summations based on the binary eccentricity. This is the same adaptation as discussed in the previous section for the Bayesian analysis.

The antenna pattern functions $\tilde{F}^A(\hat\Omega)$ are related to $F^A(\hat\Omega)$ by,
\begin{equation}
\begin{pmatrix}F^+\\F^\times\end{pmatrix} = \begin{pmatrix}\cos(2\psi) & -\sin(2\psi)\\\sin(2\psi) & \;\;\;\cos(2\psi)\end{pmatrix}\begin{pmatrix}\tilde{F}^+\\\tilde{F}^\times\end{pmatrix}.
\end{equation}
The coefficients $\mathpzc{w}_i$ are a function of {\it extrinsic} source parameters $\{\zeta,\iota,\psi,\gamma\}$, whilst the time-dependent basis-functions $\mathcal{W}^i$ are a function of {\it intrinsic} source parameters $\{F,\theta,\phi,e,l_0\}$.       

Hence, the full PTA signal template can be written as:
\begin{equation} \label{eq:signal-sum}
\mathbf{s}(t) = \sum_{i=1}^6\mathpzc{w}_i\mathcal{\mathbf{W}}^i(t),
\end{equation}
where,
\begin{equation} 
\mathcal{\mathbf{W}}^i = \begin{bmatrix}\mathcal{W}^i_1(t)\\\mathcal{W}^i_2(t)\\\vdots\\\mathcal{W}^i_{N_p}(t)\end{bmatrix},
\end{equation}
and $\mathcal{W}^i_j(t)$ denotes the quantity $\mathcal{W}^i$ defined by Eqs~(\ref{eq:Fe-basis}) for pulsar $j$. Inserting Eq.\ (\ref{eq:signal-sum}) into Eq.\ (\ref{eq:likelihood-ratio}) and using Einstein summation convention, we have
\begin{equation} 
\ln\Lambda = \mathpzc{w}_iN^i- \frac{1}{2}M^{ij}\mathpzc{w}_i\mathpzc{w}_j,
\end{equation}
where $N^i=(\delta\mathbf{t}|\mathbf{W}^i)$ and $M^{ij}=(\mathbf{W}^i|\mathbf{W}^j)$. By maximizing the log-likelihood ratio over the amplitude coefficients, $\mathpzc{w}_i$, we get their maximum-likelihood values:
\begin{equation} \label{eq:maxlike-coeff}
\hat{\mathpzc{w}}_i = M_{ij}N^j,
\end{equation}
where $M_{ij} = (M^{ij})^{-1}$.\footnote{Through practical experience we find that the inverted matrix has greater numerical stability at low eccentricity ($e\lesssim 0.05$) when a Moore-Penrose pseudoinverse is used, with a typical singular value cutoff of $\sim 10^{-10}$.} Substituting these coefficients back into the expression for $\ln\Lambda$ gives the eccentric $\mathcal{F}_e$ statistic:
\begin{equation}
\mathcal{F}_e = \frac{1}{2}N^iM_{ij}N^j.
\end{equation}

The procedure to estimate the maximum likelihood values of all of the signal parameters is as follows:
\begin{itemize}
\item We find the local maxima of the $\mathcal{F}_e$ statistic in the space of intrinsic parameters via a straightforward function maximization, or we can map out the posterior distribution of the semi-maximized parameter space with a stochastic sampler, and determine the maximum likelihood point from the resulting chain.
\item The intrinsic parameters which maximize the $\mathcal{F}_e$ statistic can be used to compute the quantities $M_{ij}$ and $N^i$, which are combined to determine the maximum likelihood coefficients, $\mathpzc{w}_i$, via Eq.\ (\ref{eq:maxlike-coeff}).
\item From these coefficients, we obtain a maximum likelihood estimate of the physical extrinsic parameters, as described below. 
\end{itemize} 
There are six $\mathpzc{w}_i$ parameters, but these are functions of only four physical extrinsic parameters and so not all combinations of $\mathpzc{w}_i$'s correspond to physical systems. However, we can obtain extrinsic parameter estimates from estimates of the $\mathpzc{w}_i$'s following \citep{cornish-porter-2007}. We define
\begin{align}
\!\!A_+\!\! =& \sqrt{(\!\mathpzc{w}_1\!\! +\!\! \mathpzc{w}_5\!)^2 + (\!\mathpzc{w}_2\!\! -\!\! \mathpzc{w}_4\!)^2} + \sqrt{(\!\mathpzc{w}_1 \!\!-\!\! \mathpzc{w}_5\!)^2 + (\!\mathpzc{w}_2 \!\!+\!\! \mathpzc{w}_4\!)^2},\nonumber\\
\!\!A_\times\!\! =& \sqrt{(\!\mathpzc{w}_1\!\! +\!\! \mathpzc{w}_5\!)^2 + (\!\mathpzc{w}_2\!\! -\!\! \mathpzc{w}_4\!)^2} - \sqrt{(\!\mathpzc{w}_1 \!\!-\!\! \mathpzc{w}_5\!)^2 + (\!\mathpzc{w}_2 \!\!+\!\! \mathpzc{w}_4\!)^2},
\end{align}
and
\begin{equation}
A = A_+ + \sqrt{A_+^2 - A_\times^2}.
\end{equation}
By employing the quantities $\{A_+,A_\times,A\}$ we can map from $\mathpzc{w}_{i\in[1,6]}$ to $\{\zeta,\iota,\psi,\gamma\}$ with the following manipulations:
\begin{align}
\zeta &= \frac{A}{4}, \nonumber\\
\cos\iota &= -\frac{A_\times}{A}, \nonumber\\
\tan(2\psi) &= \frac{A_\times\mathpzc{w}_1 - A_+\mathpzc{w}_5}{A_\times\mathpzc{w}_4 + A_+\mathpzc{w}_2}, \nonumber\\
\tan(2\gamma) &=  \frac{A_\times\mathpzc{w}_1 - A_+\mathpzc{w}_5}{A_+\mathpzc{w}_4 + A_\times\mathpzc{w}_2}.
\end{align}

In the following we treat the eccentric $\mathcal{F}_e$ statistic as a likelihood function and map out the posterior probability distribution of the semi-maximized signal parameter space. 

\begin{figure*}
   \centering \subfloat{\includegraphics[width=0.5\textwidth]{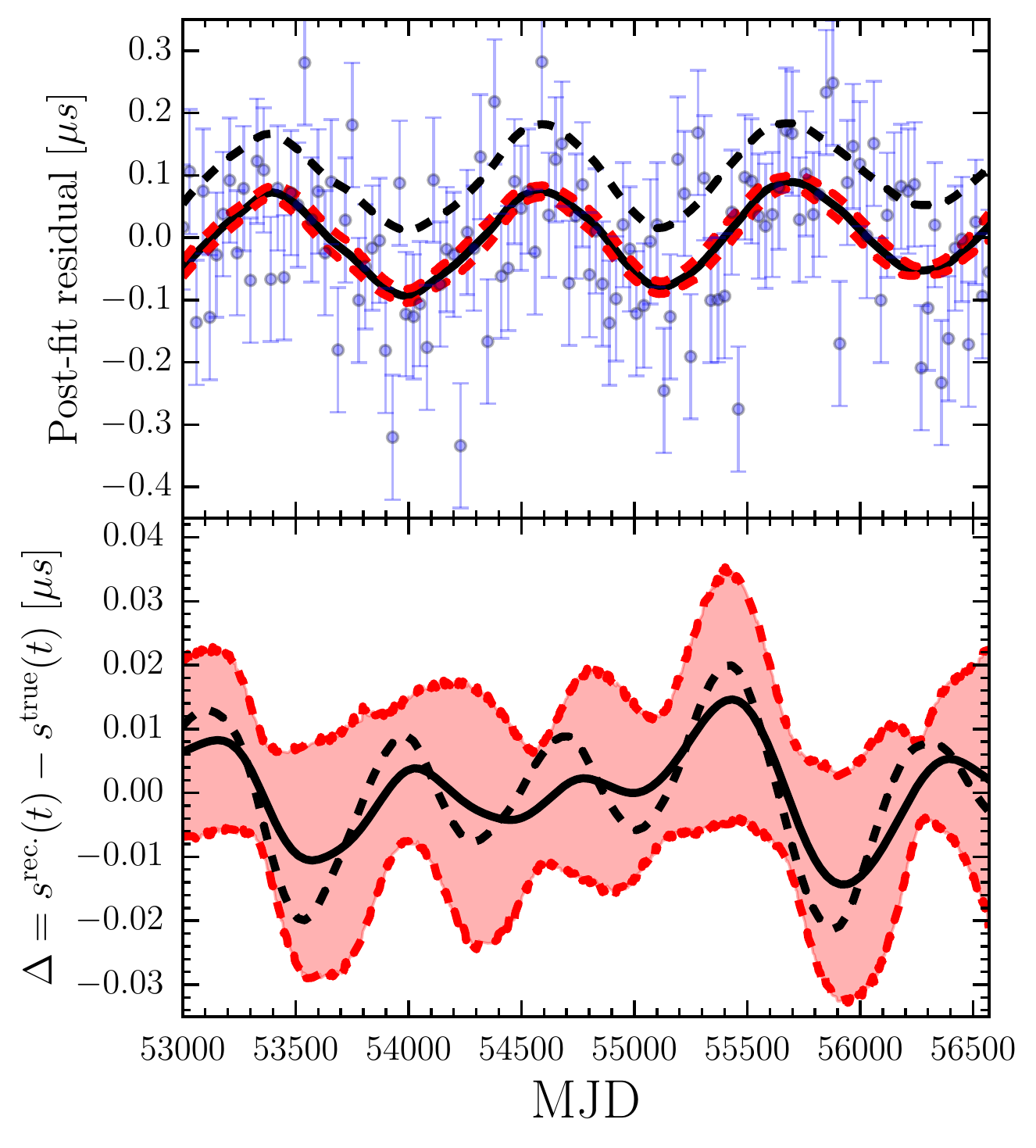}}
    \subfloat{\includegraphics[width=0.5\textwidth]{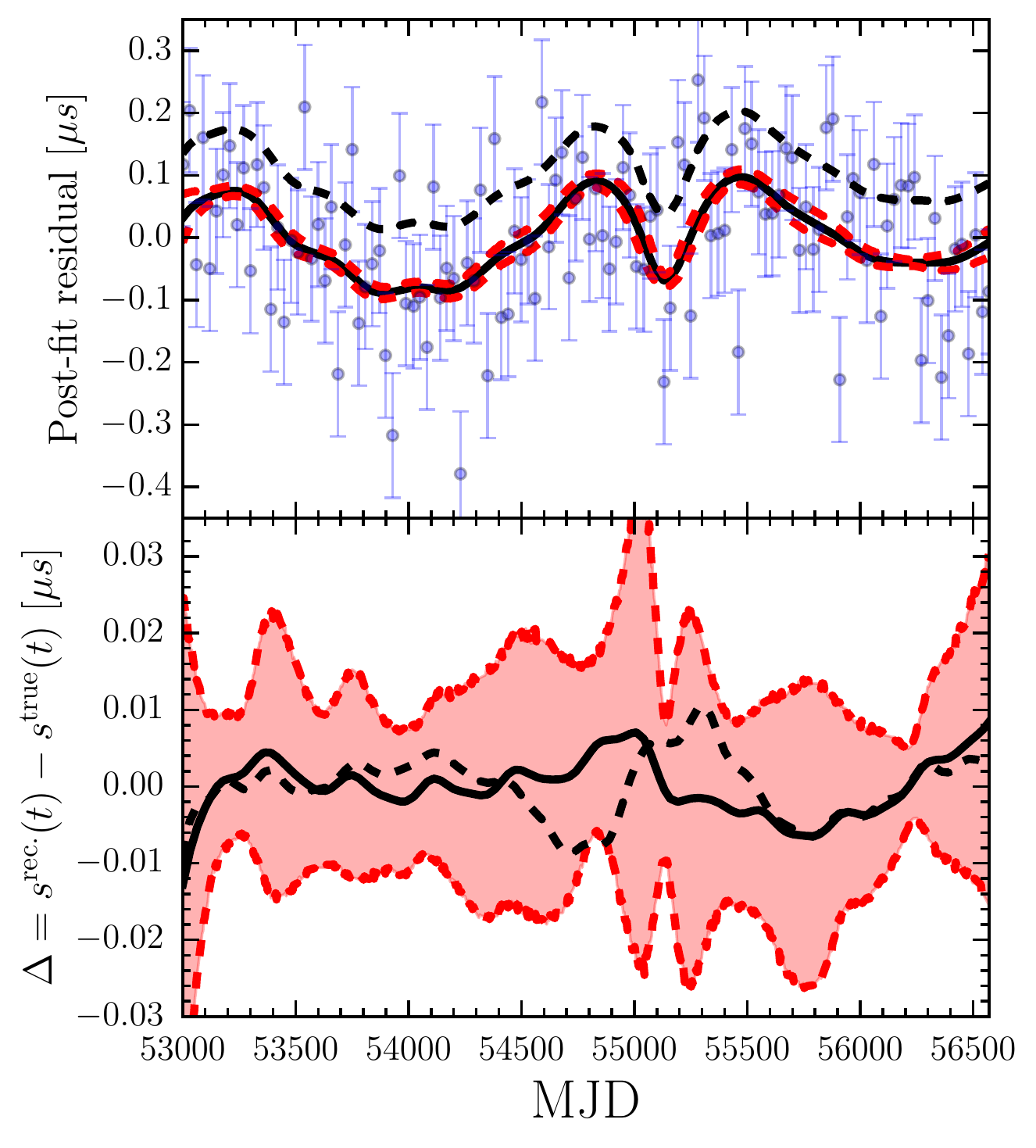}}
    \caption{The post-fit residuals of pulsar J0030+0451 for simulated \textit{Type I} data are shown in the upper portions of both panels as blue points with associated error bars. The left panel corresponds to an injected GW signal from a circular ($e=0.0$) binary, while the right panel corresponds to an injected GW signal from an $e=0.5$ binary. {\it (Upper):} The boundaries of the $95\%$ credible envelope of post-fit residuals induced by the GWs are shown as red dashed lines, while the residuals corresponding to the mean signal parameters are shown as solid black. These GW residuals are computed from the parameter posterior PDFs returned by Bayesian analysis of the simulated data, and then projected to post-fit values \citep{Demorest:2013ff}. The black dashed line shows the maximum likelihood post-fit residuals returned by an eccentric $\mathcal{F}_e$-statistic (see Sec.\ \ref{sec:eccFstat-sec}) analysis (residuals are offset by $+0.1$ $\mu$s for ease of viewing). {\it (Lower):} The offset of the reconstructed GW-induced residuals from the injected residuals is shown, where all lines correspond to the same cases as the upper panels. The boundaries of the $95\%$ Bayesian credible envelope of post-fit residuals encompasses $\Delta =0$, which is a good indicator of the robustness of the pipeline.}
    \label{fig:earth-term-waveform}
  \end{figure*}
  
\section{Results} \label{sec:results-sec}

\subsection{Efficacy of pipelines} \label{sec:bayesian-analysis}

We accelerate the generation of templates for the GW-induced residuals by making the number of waveform harmonics adapt based on the current proposed eccentricity. As discussed in Sec.\ \ref{sec:kepler-problem}, the number of harmonics to adequately describe a binary with $e=0.5$ is $\sim 10$, whilst for $e=0.9$ it is $\sim 100$. Adaptation of the number of harmonics avoids template generation being the main computational bottleneck in our pipeline.

As a first illustration of the efficacy of our pipelines, we inject GW signals with $\mathrm{SNR}=20$ into noisy \textit{Type I} datasets, and analyze the data with our Bayesian and frequentist statistics. We overlay the $95\%$ envelope of Bayesian credible post-fit GW-induced residuals on top of the raw post-fit residuals from a single pulsar in our array. The results are shown for an $e=0$ and $e=0.5$ binary signal in Fig.\ \ref{fig:earth-term-waveform}, where we see that the region of credible residuals (enclosed within red dashed lines) tracks the main features in the raw post-fit residuals, and correctly interprets high frequency behavior around MJD $55100$ in the right panel ($e=0.5$) as binary periapsis. In Fig.\ \ref{fig:earth-term-waveform} we also show the deviation of the recovered residuals from the true injected residuals, where the envelope of credible residuals encompasses the line of zero offset. This shows that, even in this high SNR case, any systematic bias from the adaptation of the number of harmonics is very small, and our Bayesian pipeline is robustly recovering the signal characteristics.

We also use our samplers to map out the $\mathcal{F}_e$ statistic distribution over the intrinsic parameter space. From the chain of sampled points we determine the maximum-likelihood intrinsic parameters, which are then used to construct $\mathpzc{w}_i$ via Eq.~(\ref{eq:maxlike-coeff}). Having the maximized $\mathpzc{w}_i$ and corresponding $\mathcal{W}^i$, we now compute the maximum-likelihood timing residuals induced by the GWs from an eccentric binary. The results for the $e=0$ and $e=0.5$ binary signals are shown in Fig.\ \ref{fig:earth-term-waveform}, where the maximum-likelihood GW-induced post-fit residuals are overlaid as black dashed lines on top of the raw post-fit residuals from pulsar J0030+0451, showing excellent tracking of the residual behavior and good agreement with the Bayesian recovery. Note that these maximum-likelihood residuals are offset by $+0.1$ $\mu$s for ease of viewing. 

\subsection{Detection prospects \& parameter precision}

\begin{figure}[!h]
\centering
\hspace{-0.18in}\includegraphics[angle=0, width=0.5\textwidth]{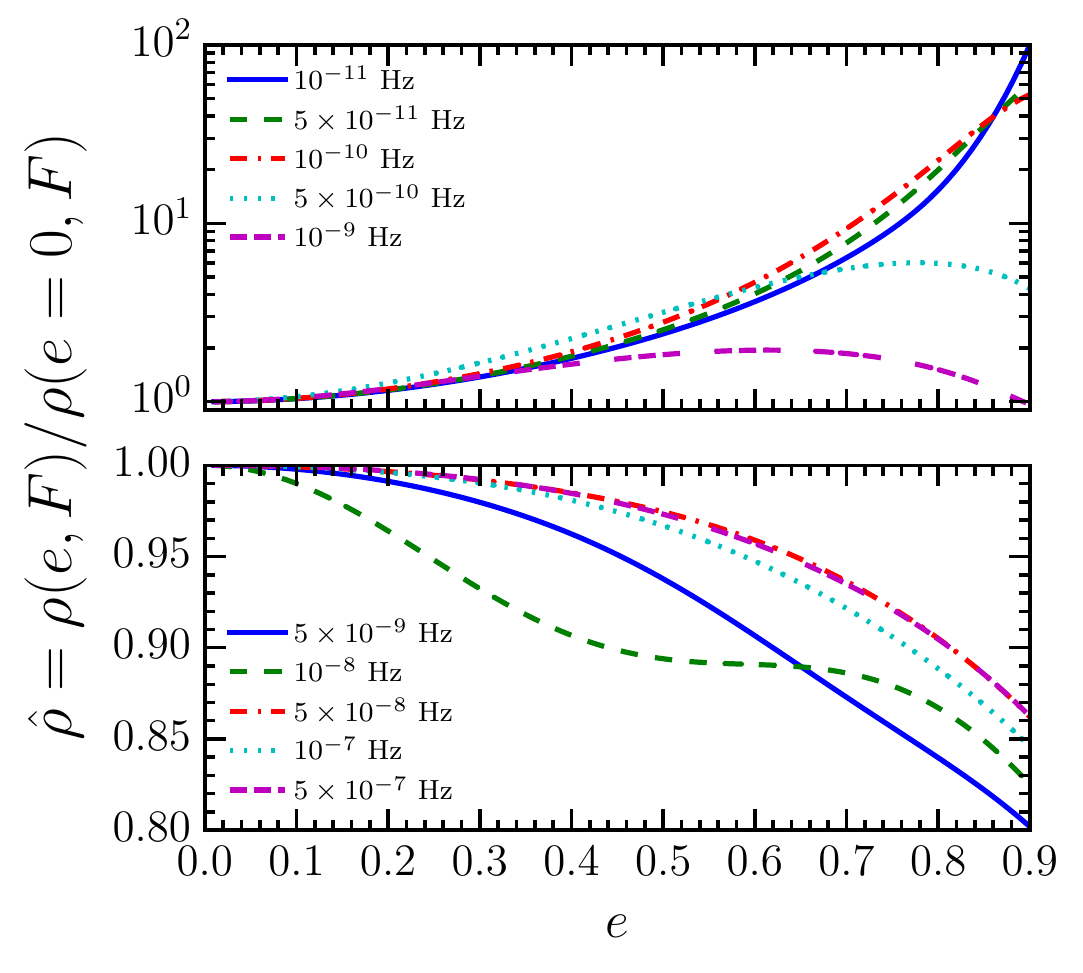}
\caption{Normalized optimal SNR of a single source as a function of the binary eccentricity for a PTA timing baseline of $10$ years (\textit{Type I} data). Only the Earth-term component is considered. Each curve corresponds to a different choice of binary orbital frequency, and is computed by averaging the SNR over all waveform angular parameters. 
For reference, the GW frequency of greatest sensitivity in this \textit{Type I} pulsar array is $\sim 5$ nHz.}
\label{fig:snr-ecc-5nhz}
\end{figure}

\begin{figure}[!ht]
\centering
\includegraphics[angle=0, width=0.5\textwidth]{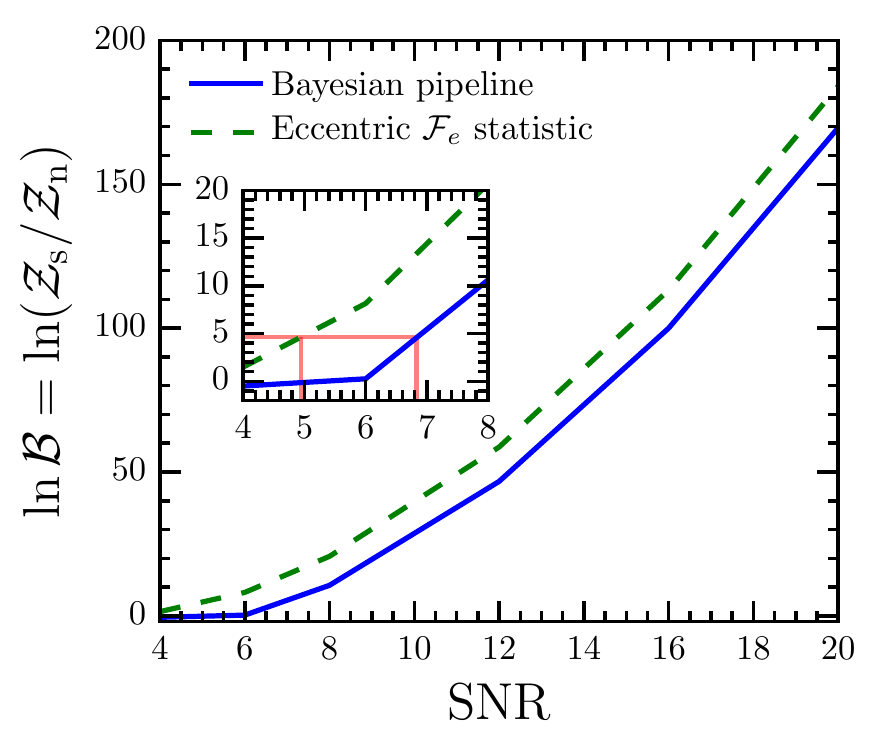}
\caption{Bayes factors for a signal+noise model versus a noise model alone in \textit{Type II} datasets with varying SNR injections. The injected binary parameters are the fiducial values given at the start of Sec.~\ref{sec:analysis-details}. The solid blue line shows the results for the full eccentric Bayesian pipeline, while the dashed green line shows the results for searches over the semi-maximized signal parameter space (intrinsic parameters) in the eccentric $\mathcal{F}_e$ statistic. Red lines in the inset figure show the SNR at which each technique reaches a Bayes factor of $100$.}
\label{fig:bayes_sigvnoise}
\end{figure}

\begin{figure*}[!ht]
\centering
\includegraphics[angle=0, width=\textwidth]{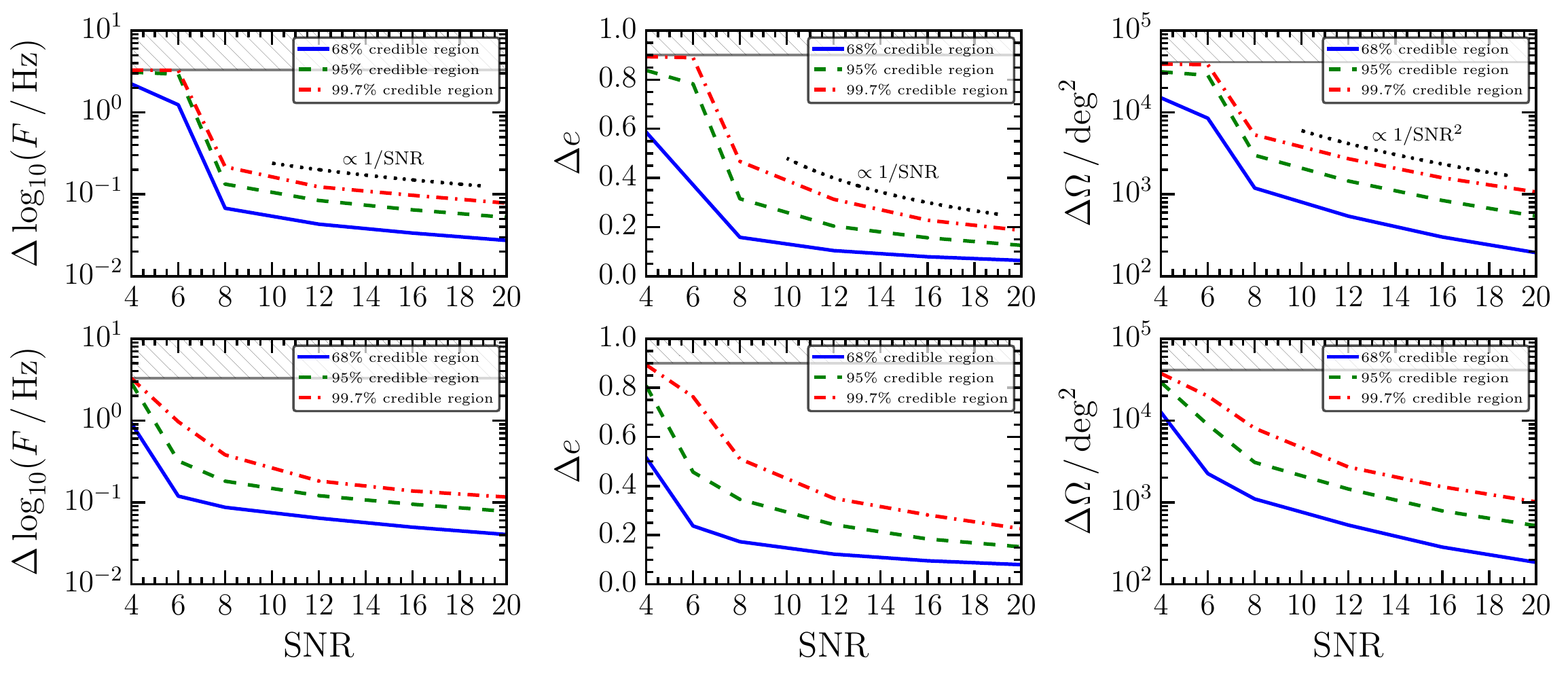}
\caption{Expected binary parameter measurement precisions from varying SNR injections into \textit{Type II} datasets. The values plotted are the full width of the respective Bayesian credible regions. The hatched regions correspond to the parameter's prior boundary. The injections are noiseless (but with pulsar noise characteristics modeled in the likelihood functions) and carried out at $F=5$ nHz, $e=0.5$. The top row corresponds to precisions deduced from the width of Bayesian credible regions produced by our Bayesian pipeline analysis. The bottom row corresponds to precisions deduced from the width of Bayesian credible regions produced by mapping the posterior distribution of the intrinsic parameter space with the eccentric $\mathcal{F}_e$ statistic.}
\label{fig:param_precision}
\end{figure*}

One might expect that distinctive high-frequency features due to periapsis passage (such as seen in Fig.\  \ref{fig:earth-term-waveform}) may improve the prospects for detection. We investigate this by computing the optimal SNR for a binary with varying orbital frequency, and a PTA timing baseline of $10$ years in \textit{Type I} data. We draw the angular waveform parameters randomly and average over the resulting SNRs. 
The result of this procedure as a function of binary eccentricity is shown in Fig.\ \ref{fig:snr-ecc-5nhz}, where we see a transition in behavior as the binary orbital frequency moves through the most sensitive location in the pulsar-timing band. From theoretical calculations and analysis of real data \citep{moore-2015,yardley2010,arz+2014}, we expect the region of peak PTA sensitivity to a continuous GW to be at a GW frequency of $\sim 1/T - 2/T$. Sensitivity is inhibited at lower frequencies by fitting of the pulsar quadratic spindown parameters in its timing-model, and higher frequencies are dominated by white TOA measurement errors. For $e=0$ binary signals in this simulated PTA, this peak corresponds to an orbital frequency of $\sim 1.6 - 3.2$ nHz.
 In Fig.\ \ref{fig:snr-ecc-5nhz} we see that at higher eccentricities the SNR is enhanced when the injected orbital frequency lies below $1$ nHz, and diminished when it lies above $5$ nHz. We can make sense of this by recalling the spectral decomposition of the variance of the GW-induced residuals shown in Fig.\ \ref{fig:PMgne-fig}, where as the eccentricity is increased the variance is distributed amongst higher harmonics of the orbital frequency. For systems with $F\lesssim 1$ nHz this will enhance the SNR since power in the residual variance is shifted into the region of peak PTA sensitivity, while for systems with $F\gtrsim 5$ nHz this diminishes the SNR since the power in the residual variance is distributed into higher, less sensitive frequencies of the PTA band.

With the approximate scaling behavior of SNR with signal eccentricity established, we now investigate how SNR maps to the Bayes factor of a signal+noise model versus a noise model alone. Bayesian model selection is actually carried out by computing the posterior odds ratio, which is the ratio of competing model evidences (Bayes factor, $\mathcal{B}$) multiplied by the prior odds ratio of each model. However, in the following we treat the latter quantity as being unity since in real searches we can not judge the \textit{a priori} odds of a signal being in our data. Assuming fixed noise properties, the computation of the Bayes factor follows by integrating the likelihood ratio in Eq.~(\ref{eq:likelihood-ratio}) over the signal parameter space. We judge a model to be favored over another if the Bayes factor exceeds $100$ ($\ln\mathcal{B}\gtrsim 4.6$). 

We inject varying SNR signals into noiseless \textit{Type II} datasets with fiducial parameters and $e=0.5$. This eccentricity is a compromise between being a moderate value in our range of exploration, and (as seen later) where the discrimination between eccentric versus circular signal models is greatest. The injections are noiseless so that we can avoid the need to perform a large program of injections to average over noise realizations \citep{nhh+10,c10}. There are some reservations over this approach in the low SNR regime \citep{v11}, however it is nevertheless correct at high SNR, so that we can consider the conclusions drawn here as optimistic but indicative of general trends. Importantly, in the following we verified that the peak of the recovered posterior distributions matched the injected signal parameters, which should be the case when the datasets are noiseless. This confirms that our techniques do not suffer as a result of the irregular sampling and heteroscedastic uncertainties associated with real data.

Figure \ref{fig:bayes_sigvnoise} shows the growth of Bayes factors favoring a signal+noise model over a noise model alone in both the full signal parameter space (Bayesian pipeline) and intrinsic parameter space (eccentric $\mathcal{F}_e$ statistic). Since the $\mathcal{F}_e$ statistic is already maximized over half of the full signal parameter space, it has a lower search dimensionality than the full Bayesian pipeline and thus receives less of an Occam penalty. As seen in the inset of Fig.\ \ref{fig:bayes_sigvnoise} the $\mathcal{F}_e$ statistic reaches a Bayesian detection threshold at $\mathrm{SNR}\sim 5$ whilst the full Bayesian pipeline does so at $\mathrm{SNR}\sim 7$. 

An important question associated with GW detection is whether a threshold signal will be associated with any meaningful parameter measurement precisions. We address this issue by analyzing the widths of the $\{68\%,95\%,99.7\%\}$ Bayesian credible regions with respect to the prior widths for binary orbital frequency, eccentricity, and sky location, at varying SNR. Our datasets are again of \textit{Type II} with fiducial parameters and $e=0.5$. The results are shown in Fig.\ \ref{fig:param_precision}, with measurement precisions obtained using the Bayesian pipeline along the top row, and precisions obtained with the eccentric $\mathcal{F}_e$ statistic along the bottom row. At high SNR the precisions for $\log_{10}F$ and $e$ obey a $1/\mathrm{SNR}$ scaling, and the sky location (being a compound of two parameters) obeys a $1/\mathrm{SNR}^2$ scaling. Both techniques perform comparably for $\mathrm{SNR}>8$, however if we look at the width of the $99.7\%$ credible region, we see that the eccentric $\mathcal{F}_e$ statistic begins to update our prior knowledge of the parameter space at $\mathrm{SNR}\gtrsim 5$, whilst this happens at $\mathrm{SNR}\gtrsim 7$ for the full Bayesian pipeline. Hence, binary parameter measurement precisions become non-trivial once the Bayes factor favoring the presence of a signal exceeds our threshold value of $100$. \citet{rsg2015} investigated the likely properties of the first detectable continuous GW source in IPTA and SKA \citep{jhm+15} data, observing that the detection probability favored massive, nearby binaries with orbital frequencies $\lesssim 10$ nHz. However, the authors did not consider eccentricity. From our results, we see that a threshold detection will provide an eccentricity measurement precision of $\sim 0.3$ (considering the $68\%$ credible region width), which may allow the eccentricity to be sufficiently constrained as to perform inference on plausible environmental coupling influences (such as 3-body stellar scattering or circumbinary-disk interaction) which drove the binary's orbital evolution. Doing so will shed light on the astrophysics and environment of the binary's host galactic nucleus.

\subsection{Circular-model penalty}

We now address the detection penalty one might incur by searching for an eccentric binary signal with a circular waveform model, which was also investigated in \citet{zhu2015}
using frequentist methods. 
We firstly investigate this using signal injections in \textit{Type I} data for a variety of injected binary frequencies and eccentricities. The matched-filtering SNR for a circular (monochromatic) template in data with an eccentric signal is computed, and compared to the optimal SNR of the same eccentric signal. The resulting statistic, $\rho_\mathrm{circ}/\rho_\mathrm{opt}$, is a measure of the \textit{effectualness} of the circular template in representing the eccentric signal \citep{pnbuo}. At each eccentricity, the SNR is averaged over $10^3$ binary orientations and locations, and maximized over the frequency of a monochromatic template. The matched-filtering SNR is computed in three different ways: $(a)$ as a coherent SNR for the entire pulsar array, maximized over the monochromatic template frequency; $(b)$ as a coincident SNR, with the SNR in each pulsar independently maximized over the monochromatic template frequency, and then added in quadrature to give the full array statistic; $(c)$ as a coincident SNR, with the SNRs added in quadrature to give the full array statistic, but demanding a common frequency for the monochromatic template.

\begin{figure}[!ht]
\centering
\hspace{-0.18in}\includegraphics[angle=0, width=0.5\textwidth]{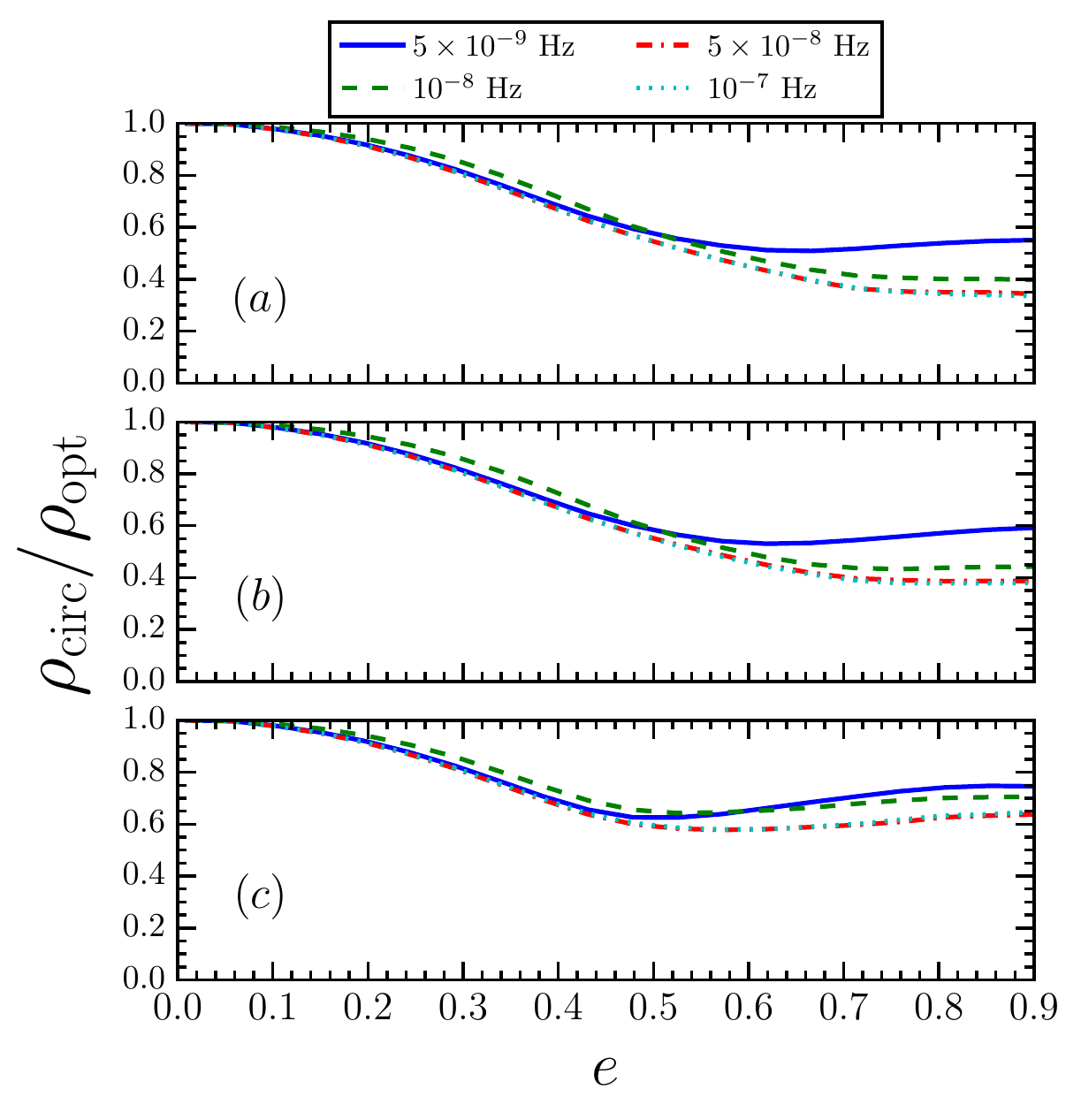}
\caption{Ratio of circular-template matched-filtering SNR to optimal SNR for signals with various orbital frequencies and eccentricities. Case $(a)$ shows results for a coherent array SNR. Case $(b)$ shows results for a coincident array SNR with independently maximized template frequencies in each pulsar. Case $(c)$ shows results for a coincident array SNR with a common monochromatic template frequency. Further details and discussion are provided in the text.}
\label{fig:circ-temp-penalty}
\end{figure}

Our results for \textit{Type I} data are shown in Fig.\ \ref{fig:circ-temp-penalty} for orbital frequencies beyond the region of peak PTA sensitivity ($\gtrsim 5$ nHz). For cases $(a)$ and $(b)$, the favored monochromatic template frequency is twice the orbital frequency until $e\sim 0.5 - 0.6$, and incurs an increasingly harsh SNR penalty as the eccentricity of the signal is increased. However, beyond $e\sim 0.5 - 0.6$ the SNR recovers slightly, since the template frequency now favors the fundamental harmonic of the signal, which is lower and closer to the region of peak PTA sensitivity. This is seen even more clearly in case $(c)$, where there is a common monochromatic template frequency across all pulsars when constructing the coincident SNR. The loss in SNR is slightly greater in $(a)$ than in $(b)$ and $(c)$, since in the former we require signal coherence amongst all pulsars in the array. The behaviour found for these three cases is likely pessimistic, since in real matched-filtering searches the SNR is maximized over all template parameters rather than just the frequency. Lower orbital frequency SNR curves exhibit similarly increasing penalties as the eccentricity is raised, but the trends are not as smooth. 

\begin{figure}[!ht]
\centering
\hspace{-0.18in}\includegraphics[angle=0, width=0.5\textwidth]{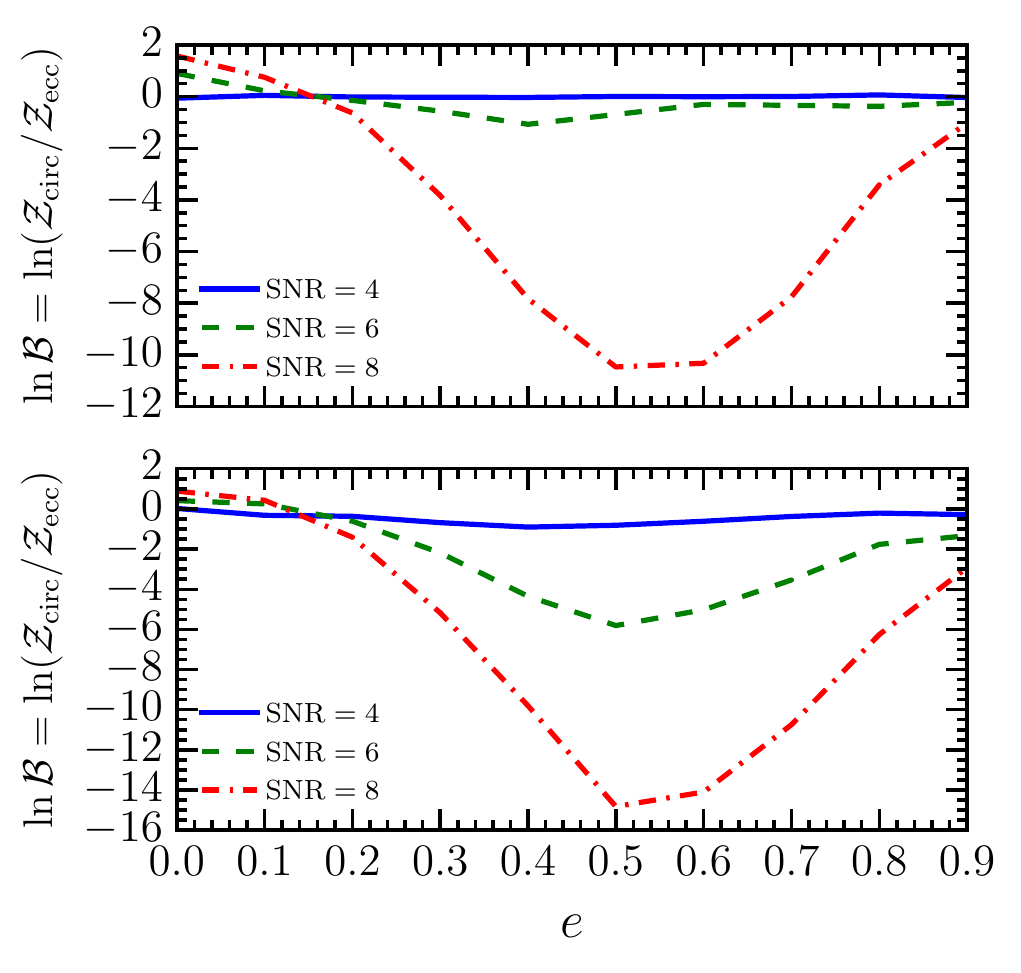}
\caption{Bayes factors, $\mathcal{B}$, for circular versus eccentric signal models when analyzing \textit{Type II} datasets which have signals with varying injected eccentricity. The upper panel shows results for the full Bayesian analysis, while the lower panel shows results from mapping the posterior distribution of the intrinsic parameter space with the eccentric $\mathcal{F}_e$ statistic.}
\label{fig:bayes_circpenalty}
\end{figure}

We now investigate the circular-model penalty in terms of Bayesian model selection, by injecting fiducial signals of varying eccentricity and SNR into \textit{Type II} data. The results are shown in Fig.\ \ref{fig:bayes_circpenalty}, with the quoted Bayes factors corresponding to circular versus eccentric signal models. The top panel shows the results for the full Bayesian pipeline, whilst the bottom panel shows results for the eccentric $\mathcal{F}_e$ statistic. Both techniques exhibit the same general trends: $(1)$ at eccentricities $\lesssim 0.1$ the eccentric model receives an Occam penalty, resulting in the circular model being slightly favored, although not decisively so; $(2)$ as the signal eccentricity increases so does the circular-model penalty, until the models are most easily discriminated at $e\sim 0.5 - 0.6$; $(3)$ at higher eccentricities the signal model is being dominated by the fundamental harmonic, allowing the circular model to function as a better approximation to the injected signal than at intermediate eccentricities, resulting in a reduction in the circular-model penalty. These trends are qualitatively similar to those found by \citet{zhu2015} in Fig.~(13) of their paper, however our eccentric search strategies exhibit superior performance at high eccentricity by virtue of modeling the distribution of signal harmonics with in-code adaptation (see Fig.~\ref{fig:minharms}) rather than just including the lowest two harmonics. A key result of our analysis is that we require $\mathrm{SNR}>8$ in order for an eccentric signal model to be correctly discriminated and favored when the true eccentricity is greater than $\sim 0.3$.

\section{Caveats $\&$ future directions} \label{sec:caveats-future}

The analysis and results presented in this paper have relied on several assumptions. We discuss these here, and the prospects for relaxing these caveats in future work.

\subsection{Prospects for including the pulsar term} 

\begingroup
\setlength{\tabcolsep}{3pt}
\begin{table*} 
\caption{\raggedright The matched-filter SNR for an Earth term template is compared against the optimal full signal SNR to construct $\rho_\mathrm{earth}/\rho_\mathrm{full}$. At each Earth term orbital frequency and eccentricity, we evolve a binary backwards in time by $L(1+\hat\Omega\cdot\hat{u})$ to construct the pulsar term waveform, where $L=1$ kpc for all pulsars.}
\label{tab:earth-suboptimal}
\centering
\begin{tabular*}{1.0\textwidth}{@{\extracolsep{\fill}} c | c c c c  | c c c c | c c c c | c c c c | c c c c}
\hline
\multirow{4}{*}{Orbital frequency [nHz]} & \multicolumn{20}{c}{Eccentricity} \\
& \multicolumn{4}{c}{$e=0.0$} & \multicolumn{4}{c}{$e=0.25$} & \multicolumn{4}{c}{$e=0.50$} & \multicolumn{4}{c}{$e=0.75$} & \multicolumn{4}{c}{$e=0.90$} \\\noalign{\vskip 1mm}    
\cline{2-21}
& \multicolumn{4}{c|}{$\mathcal{M}\;[M_\odot]$} & \multicolumn{4}{c|}{$\mathcal{M}\;[M_\odot]$} & \multicolumn{4}{c|}{$\mathcal{M}\;[M_\odot]$} & \multicolumn{4}{c|}{$\mathcal{M}\;[M_\odot]$} & \multicolumn{4}{c}{$\mathcal{M}\;[M_\odot]$} \\
& $10^{7}$ & $10^{8}$ & $10^{9}$ & $10^{10}$ & $10^{7}$ & $10^{8}$ & $10^{9}$ & $10^{10}$ & $10^{7}$ & $10^{8}$ & $10^{9}$ & $10^{10}$ & $10^{7}$ & $10^{8}$ & $10^{9}$ & $10^{10}$& $10^{7}$ & $10^{8}$ & $10^{9}$ & $10^{10}$ \\
\hline
\hline
$0.1$ & 0.76 & 0.76 & 0.76 & 0.76 & 0.73 & 0.73 & 0.73 & 0.73 & 0.51 & 0.51 & 0.51 & 0.51 & 0.12 & 0.12 & 0.12 & 0.12 & 0.13 & 0.13 & 0.13 & 0.22\\
$0.5$ & 0.78 & 0.78 & 0.78 & 0.78 & 0.66 & 0.66 & 0.66 & 0.66 & 0.33 & 0.33 & 0.33 & 0.34 & 0.15 & 0.15 & 0.16 & 0.41 & 0.07 & 0.07 & 0.36 & 0.87\\
$1.0$ & 0.63 & 0.63 & 0.63 & 0.64 & 0.47 & 0.47 & 0.47 & 0.49 & 0.25 & 0.25 & 0.25 & 0.36 & 0.10 & 0.10 & 0.16 & 0.76 & 0.05 & 0.12 & 0.69 & 0.96\\
$5.0$ & -0.03 & -0.03 & 0.07 & 0.65 & -0.02 & -0.02 & 0.13 & 0.72 & 0.0 & 0.01 & 0.38 & 0.87 & 0.02 & 0.16 & 0.81 & 0.99 & 0.11 & 0.71 & 0.99 & 1.0\\
$10.0$ & -0.01 & 0.02 & 0.59 & 0.62 & -0.03 & 0.01 & 0.6 & 0.73 & -0.04 & 0.10 & 0.64 & 0.93 & 0.03 & 0.50 & 0.87 & 1.0 & 0.34 & 0.81 & 0.99 & 1.0\\
$50.0$ & 0.21 & 0.68 & 0.63 & 0.42 & 0.3 & 0.68 & 0.56 & 0.95 & 0.53 & 0.67 & 0.81 & 0.99 & 0.68 & 0.66 & 0.99 & 1.0 & 0.60 & 0.97 & 1.0 & 1.0\\
$100.0$ & 0.66 & 0.65 & 0.55 & 0.34 & 0.67 & 0.64 & 0.65 & 0.99 & 0.68 & 0.52 & 0.93 & 1.0 & 0.64 & 0.81 & 0.99 & 1.0 & 0.73 & 0.99 & 1.0 & 1.0\\
\hline\hline
\end{tabular*}
\end{table*}
\endgroup

In the majority of this paper, we have ignored a full treatment of the pulsar term signal. Since the pulsar term is retarded with respect to the Earth term, it will represent the binary at an earlier stage of its orbital evolution, with a larger eccentricity and smaller orbital frequency. It is now well known that the pulsar term aids detection prospects for continuous wave sources, and is crucial in breaking degeneracies between the binary mass and its luminosity distance by providing extra information from the binary's evolution over the lag time between the Earth and pulsar term signals \citep{corbin-cornish-2010,lee+2011,ellis2013}.

Being able to model the orbital evolution of the binary, and constrain the properties of this evolution through continuous GW searches with PTAs, will provide a unique opportunity to probe the influence of other non-GW driving mechanisms. For example, the rate at which the binary orbital frequency, $F$, is driven by GWs, stellar scattering, and circumbinary disk interactions, scales as $\propto F^{11/3}$, $\propto F^{1/3}$, $\propto F^{4/3}$, respectively \citep{Sesana:2013CQG}. If we can include parametrized models of the rate of binary evolution in constructing full Earth and pulsar term signal models in a Bayesian or frequentist search, then we will be able to make statements about the relative importance of the aforementioned mechanisms. This in itself may provide clues as to how binaries are driven to sub-parsec orbital separations after dynamical friction in post-merger galaxies becomes inefficient, thereby adding to our knowledge of how the final parsec problem \citep{milomerritt2003} is ameliorated.

For now, we estimate the degree to which employing only the Earth term in searches is sub-optimal for detection. We compute the matched-filter SNR for an Earth term template applied to a full signal (including the pulsar term), and compare this to the optimal SNR for the full signal. In constructing the pulsar term component of the signal, we evolve the orbital parameters of the binary backwards in time according to Eq.\ (\ref{eq:dfbydt}) and the procedure outline in Sec.\ \ref{sec:timing-residual-sec}, where we assume all pulsars lie at a distance of $1$ kpc from the Earth. We assume all orbital evolution is GW driven. No pericenter-direction evolution or orbital-plane precession is considered, and we do not evolve the binary during the pulsar observation timespan of $10$ years. To ease the computational burden, we use a sub-array of $6$ pulsars spread across the sky, averaging the SNR over $10^3$ binary locations and orientations. The  observational cadences and timing baselines of the sub-array are of \textit{Type I} variety.

The results are shown in Table \ref{tab:earth-suboptimal} for a variety of Earth term orbital frequencies and eccentricities. As the orbital frequency and chirp mass are increased, the ratio of the Earth term SNR to the full SNR tends to grow with eccentricity. This is because higher mass, frequency, and eccentricity binaries are driven rapidly via GW emission, which in the most extreme cases leads to signals with pulsar term frequencies which are so far below the PTA sensitivity band that an Earth term template becomes an excellent approximation to the full signal. Even at fixed orbital frequency and eccentricity, the effect of increasing binary chirp mass is to raise the efficacy of an Earth term only template. However, care must be taken in the intermediate case, when we have moderate eccentricities, frequencies, and masses, which generate pulsar term signals that remain in the PTA band, and whose spectrum of GW frequencies may exceed the fundamental harmonic of the Earth term signal. The worst matches between signal and template occur for low mass, low eccentricity systems with orbital frequencies close to the region of peak PTA sensitvity ($\sim 5$ nHz) -- the combination of high array sensitivity and negligible orbital evolution leads to pulsar term signals close to this region of peak sensitivity, and thus very poor matches (which are sometimes negative since we employ a coherent SNR). In general, the pulsar term increases the signal detection prospects, but confusion may arise between different harmonics in the Earth and pulsar terms, which would harm parameter estimation efforts. So long as the Earth and pulsar terms remain distinguishable, we will learn more about the system parameters from the pulsar term's inclusion. Future work should study the prospects for incorporating the pulsar term in eccentric binary search strategies, and investigate the rich science that can be mined from having access to snapshots of the binary evolution from thousands of years in its past.

\subsection{Binary orbital evolution during the observation timespan}

\begin{figure}[!t]
\centering
   \includegraphics[angle=0, width=0.5\textwidth]{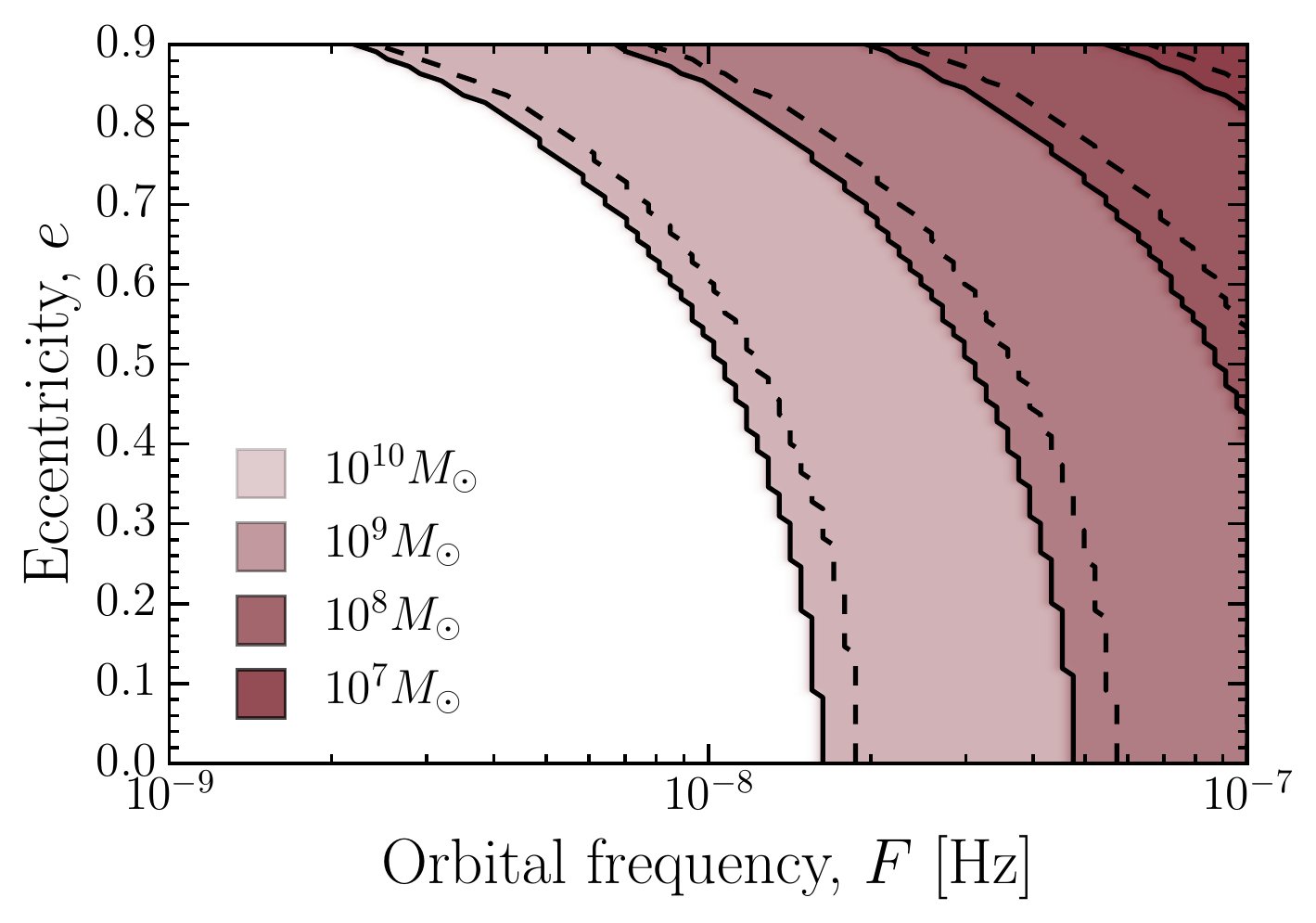}
    \caption{Exclusion regions in binary eccentricity and orbital frequency as a function of \textit{chirp mass}, corresponding to parameter combinations where the fundamental (dashed black lines) and second harmonic (solid black lines on the boundary of shaded exclusion regions) of the orbital frequency evolve during $T=10$ years by more than the PTA frequency resolution, $\Delta f=1/T=3.2$ nHz, rendering the assumption of binary non-evolution invalid.}
    \label{fig:exclusion-plots}
  \end{figure}

We now test the assumption of binary non-evolution over typical PTA observation timespans. For different initial Earth term parameter choices $\{\mathcal{M},F,e\}$, we numerically evolve a binary forward in time by $10$ years according to Eq.\ (\ref{eq:dfbydt}). Figure \ref{fig:exclusion-plots} shows exclusion regions in parameter space where the fundamental and second harmonic of the orbital frequency evolve by more than the PTA frequency resolution, ${\Delta f=1/T=3.2}$ nHz, which may render the approximation of binary non-evolution within our observing window invalid. The second harmonic will dominate the signal for low eccentricities whilst the fundamental harmonic will dominate at higher eccentricites.

A more rigorous way of testing this is to investigate how this assumption affects our ability to perform parameter estimation. If the non-evolution model performs well within the range of expected SNR, such that the systematic bias introduced via our assumption of binary non-evolution is smaller than statistical errors, then we can judge the model to be an excellent functioning approximation. More formally, we want to satisfy the \textit{indistinguishability criterion} \citep{Cutler:2007uf,creighton2012gravitational}:
\begin{equation} \label{eq:sys-inequality}
(\delta s(t)|\delta s(t)) < 1,
\end{equation}
where $\delta s(t)$ corresponds to the difference between the approximated residuals in the non-evolution model and the true residuals. Satisfying the inequality in Eq.\ (\ref{eq:sys-inequality}) approximately corresponds to the systematic errors arising from modeling bias being smaller than statistical measurement errors. The tolerance SNR, $\rho_\mathrm{tol.}$, above which systematic errors from insufficient template accuracy may exceed statistical measurement errors, and thus become problematic, is given by
\begin{equation}
\rho^2_\mathrm{tol.} = \frac {(\mathbf{s}(t)|\mathbf{s}(t))}{(\delta \mathbf{s}(t)|\delta \mathbf{s}(t))},
\end{equation}
where $\mathbf{s}(t)$ are the true residuals (concatenated over all pulsars) induced by a binary which may be evolving over our observation timespan. To compute this, we numerically evolve the orbital parameters of a binary over the $10$ year timing baseline of a \textit{Type I} dataset using Eq.\ (\ref{eq:dfbydt}), with varying choices of initial orbital frequency and eccentricity. The evolved orbit is then used to compute the pulse redshift and (via numerical integration) the GW-induced timing residual at each pulse TOA. The typical ratio of the time required to compute the GW signal numerically versus analytically is $\sim \mathcal{O}(10^4)$, which is why a fully numerical approach is clearly intractable at present. 

\begin{figure}[!t]
\centering
   \hspace{-0.18in}\includegraphics[angle=0, width=0.5\textwidth]{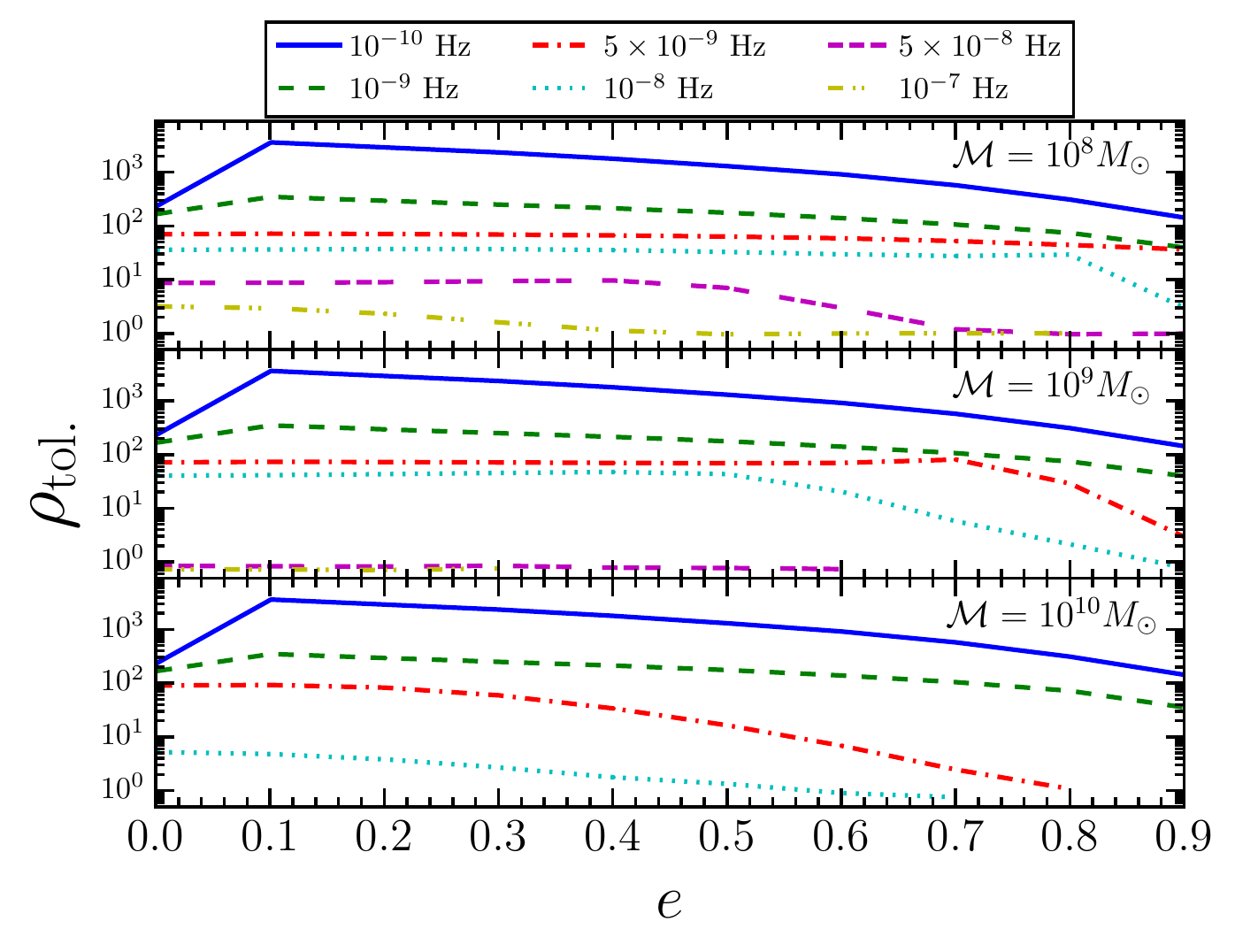}
    \caption{The tolerance SNR, $\rho_\mathrm{tol.}$, for a range of binary eccentricities, orbital frequencies, and chirp masses is shown. This indicates the SNR above which systematic parameter errors (which occur by keeping binary parameters fixed over the $10$ year PTA timing baseline) may exceed statistical measurement errors.}
    \label{fig:evolving-snrtol}
  \end{figure}

The tolerance SNR is shown in Fig.\ \ref{fig:evolving-snrtol} as a function of binary eccentricity, orbital frequency, and chirp mass. We choose a cutoff value of the tolerance SNR equal to $10$ since this may correspond to realistic values of the SNRs of first PTA detections of single GW sources after $\sim 10$ years of IPTA and SKA1 activity \citep{rsg2015}. If our model can be successfully applied to real signals above this cutoff value, then we conclude that the treatment used in this paper is valid well into the era of first PTA detections. We see that at a binary chirp mass of $10^8M_\odot$ the tolerance SNR is above $10$ for most frequencies and eccentricities, indicating that the assumption of non-evolution is valid. The approximation begins to break down at higher eccentricities and frequencies ($\gtrsim 5\times 10^{-8}$ Hz) where the rate of binary evolution is higher. At $10^9M_\odot$ our model is appropriate at all eccentricities for frequencies lower than $10^{-8}$ Hz, however the tolerance SNR for $F=10^{-8}$ Hz drops below cutoff at $e\sim 0.7$, and at higher frequencies the assumption of binary non-evolution is inappropriate. Finally, for the most massive binaries with $\mathcal{M}=10^{10}M_\odot$, the tolerance SNR remains above cutoff for orbital frequencies lower than $5\times 10^{-9}$ Hz at all eccentricities, while at $5\times 10^{-9}$ Hz the tolerance SNR only drops below $10$ at $e\sim 0.6$. 

Therefore, our assumption (which has been shared by all other authors in this field) of binary non-evolution over typical PTA timing baselines is appropriate for most frequencies at or below the region of peak PTA sensitivity. The approximation only begins to break down for the most massive systems above orbital frequencies of $\sim 5\times 10^{-9}$ Hz and eccentricities of $0.6$, allowing the signal model and analysis techniques developed in this article to be applied to real data with robust outcomes. Future studies are required to investigate faster and more tractable strategies for modeling the orbital evolution of high mass, high frequency, and high eccentricity binaries over PTA timing baselines.

\section{Conclusions} \label{sec:conclusion-sec}

PTAs are uniquely suited to explore the dynamical evolution of SMBHBs before and after they decouple from their astrophysical environments to become dominated by GW emission. An increasing number of studies tend to suggest that the mechanisms that may drive SMBHBs to small orbital separations could also lead to an increase in binary eccentricity that will be detectable in the frequency band of PTAs. Extracting this information from real data will substantially increase our understanding of the mechanisms that lead to the formation, hardening and eventual coalescence of SMBHBs. In this article we have introduced several tools to address this issue. We have developed a robust, accurate and computationally efficient Bayesian pipeline to explore the feasibility of detecting and reconstructing the astrophysical parameters of eccentric SMBHBs in PTA data, and have developed for the first time an eccentric $\mathcal{F}_e$-statistic that is, by construction, suitable to study systems of arbitrary eccentricity. 

We have used these tools to determine the accuracy with which a simulated eccentric signal could be reconstructed, and have conclusively shown that the recovered and injected parameters are completely consistent. Our prior knowledge of the eccentric binary parameter space will begin to be updated by data once the SNR of the associated binary's GW signal exceeds $\sim 7$. We have also shown that the automated waveform generation algorithm, which determines the number of harmonics needed to ensure that the modeled GW signal reproduces the full numerical solution with an accuracy better than $99.999\%$, prevents computational inefficiencies in the pipeline. 

The influence of binary eccentricity on PTA single-source detection prospects was also considered. Assuming that the sensitivity peak of a PTA to continuous wave sources is located at a GW frequency $f_0$, we have shown that eccentricity will enhance the detection prospects of SMBHBs with orbital frequencies $\lesssim f_0$. This is because the signal spectrum of eccentric binaries is distributed into higher harmonics of the orbital frequency than in the case of a circular binary, leading to components of the signal being located in the region of maximum PTA sensitivity. On the other hand, binaries with orbital frequencies $\gtrsim f_0$ will undergo an SNR attenuation because the signal power is shifted to higher frequencies where the PTA sensitivity is poorer and dominated by TOA measurement errors. In summary, systems with signals which are below band in the circular case get pushed into band through increasing eccentricity, while systems that are optimally located in frequency for the circular case get pushed out of band by eccentricity.

We found that applying a circular waveform model in the analysis of data with increasingly eccentric binary signals incurs an SNR penalty which grows with eccentricity, and is $\sim 60\%$ at worst case for coherent and coincident analyses. This was also investigated in a Bayesian context, where we found that SNRs greater than $8$ are needed in order for an eccentric signal model to be correctly discriminated and favored over a circular signal model when the true signal eccentricity is $\geq 0.3$.

Several of the approximations used in the techniques presented in this article were briefly investigated. We found that for very high mass, frequency and eccentricity binaries, an Earth term signal model performs just as well as a full signal model incorporating the pulsar term, since the binary will have evolved so significantly that the pulsar term signal lies below band. Furthermore, the possible bias from assuming binary non-evolution over a PTA observation time of $10$ years was studied, and was found to be unimportant for moderately massive and eccentric systems in the era of first PTA detections. 

There are several topics in continuous GW searches that should be addressed in the near future, including the need to assess the possible covariances involved in simultaneous continuous GW searches and stochastic GW background searches. One can imagine that the reduction in low-frequency sensitivity associated with having fit for the pulsar quadratic spindown parameters will be exacerbated by the concurrent search for a stochastic GW background signal dominated by low-frequency power. Also, the possibility of having multiple resolvable continuous-wave sources may lead to difficulties in isolating each source during the Bayesian searches, resulting in interesting covariances. However there are ongoing efforts to resolve this issue \citep{ellis15}. Furthermore, as new data is added to each PTA and combined to form IPTA datasets, the prospects for continuous GW source detection grow stronger. Current pipelines should be tested against signals injected into near-future type datasets as a means to inform new advances in analysis procedures. Finally, the efficacy of performing continuous GW searches on datasets having signals composed of realistic GW source populations must be addressed in the near future. These areas of future study have not yet been investigated with circular-binary GW signal models, however our development in this paper of more complete signals models which include eccentricity will endow these studies with greater verisimilitude.

\citet{hmgt2015} and this article have provided a solid foundation to explore in a consistent way the influence of eccentricity on the detection and parameter estimation of SMBHBs with PTAs. The tools presented in this article can be readily incorporated into all present and planned analysis pipelines. The toolkit introduced in these articles could be extended to explore in detail what constraints may be placed on the various astrophysical mechanisms that can drive the dynamical evolution of SMBHBs prior to becoming dominated by GW emission. 

\begin{acknowledgements}
We thank the anonymous referee for their remarks, which significantly improved the depth and quality of this manuscript. This research was in part supported by SRT's appointment to the NASA Postdoctoral Program at the Jet Propulsion Laboratory, administered by Oak Ridge Associated Universities through a contract with NASA. JRG's work is supported by the Royal Society. We thank Justin Ellis for useful feedback on this manuscript, and Alberto Sesana for fruitful discussions. We also thank the anonymous referee for many helpful suggestions which significantly improved this manuscript. This work was supported in part by National Science Foundation Grant No. PHYS-1066293 and the hospitality of the Aspen Center for Physics. We are grateful for computational resources provided by the Leonard E Parker Center for Gravitation, Cosmology and Astrophysics at University of Wisconsin-Milwaukee. Copyright \copyright\, 2015. All rights reserved.
\end{acknowledgements}

\bibliographystyle{apj}
\bibliography{references}

\end{document}